\documentclass[a4paper,11pt]{article}
\pdfoutput=1 
\usepackage{jheppub}

\usepackage[utf8]{inputenc}

\usepackage{float}

\usepackage{amssymb,amsmath,amsfonts}
\usepackage{mathtools}
\usepackage{mathrsfs}
\usepackage{bbm}
\usepackage{slashed}
\usepackage{nicefrac}

\usepackage{graphicx}
\usepackage[dvipsnames]{xcolor}
\usepackage{array}

\usepackage{simplewick}

\usepackage{hyperref}
\usepackage{xparse}
\usepackage{xspace}

\usepackage{tikz}
\usetikzlibrary{decorations.pathmorphing}
\usetikzlibrary{automata,positioning}

\usepackage{cancel}
\usepackage[normalem]{ulem}

\usepackage{xifthen}
\usepackage{dsfont}
\usepackage[titletoc]{appendix}
\usepackage{booktabs}
\usepackage{units}

\newcommand{\gettitle}{}
\hypersetup{linkcolor=black
	colorlinks,
	linkcolor={red!75!black},
	citecolor={blue!75!black},
	urlcolor={blue!75!black},
	pdftitle={\gettitle},
	pdfauthor={Boussarie, Mehtar-Tani},
	pdfkeywords={Perturbative QCD} {Small-x}{TMD},
	bookmarksopen=true,
	bookmarksopenlevel=2,
	bookmarksnumbered=true
}
\makeatletter
\newcommand\makebig[2]{%
  \@xp\newcommand\@xp*\csname#1\endcsname{\bBigg@{#2}}%
  \@xp\newcommand\@xp*\csname#1l\endcsname{\@xp\mathopen\csname#1\endcsname}%
  \@xp\newcommand\@xp*\csname#1r\endcsname{\@xp\mathclose\csname#1\endcsname}%
}
\makeatother
\makebig{biggg} {1.3}

\def\bs{\boldsymbol} 
\def\del{\partial}
\def\bdel{\bs\partial}

\newcommand{\eqn}[1]{Eq.~\eqref{#1}}

\long\def\comment#1{ }

\newcommand{\nn}{\nonumber\\ }

\def\be{\begin{eqnarray*}}
\def\ee{\end{eqnarray*}}
\def\beq{\begin{eqnarray}}
\def\eeq{\end{eqnarray}}
\newcommand{\bea}{\beq \begin{aligned}}
\newcommand{\eea}{\end{aligned}\eeq}

\def\k{{\boldsymbol k}}

\def\r{{\boldsymbol r}}
\def\A{{\boldsymbol A}}
\def\x{{\boldsymbol x}}
\def\y{{\boldsymbol y}}
\def\z{{\boldsymbol z}}

\def\0{{\boldsymbol 0}}

\def\k{{\boldsymbol k}}

\def\x{{\boldsymbol x}}
\def\y{{\boldsymbol y}}

\def\D{{\boldsymbol D}}
\def\r{{\boldsymbol r}}
\def\z{{\boldsymbol z}}

\def\b{{\boldsymbol b}}

\def\A{{\boldsymbol A}}

\def\A{{\boldsymbol A}}
\def\hatA{{\boldsymbol {\hat A}}}

\def\bt{\boldsymbol {\rm t}}

\def\rme{{\rm e}}
\def\rmd{{\rm d}}

\def\and{ \quad\text{and}\quad}

\def\cP{{\cal P}}

\def\cO{{\cal O}}

\usepackage{color}
\definecolor{rbcolor}{rgb}{0.7,0.1,0}

\newcommand\rbout{\marginpar{\color{rbcolor}$\clubsuit$}\bgroup\markoverwith{\color{rbcolor}{\rule[0.4ex]{2pt}{0.8pt}}}\ULon}

\definecolor{ymtcolor}{rgb}{0.1,0,0.7}

\newcommand\ymtout{\marginpar{\color{ymtcolor}$\clubsuit$}\bgroup\markoverwith{\color{ymtcolor}{\rule[0.4ex]{2pt}{0.8pt}}}\ULon}

\begin{document}

\title{On gauge invariance of transverse momentum dependent distributions at small x}

\author[a]{Renaud Boussarie} 
\author[a]{Yacine Mehtar-Tani}

\affiliation[a]{Physics Department, Brookhaven National Laboratory, Upton, NY 11973, USA}
\emailAdd{rboussarie@bnl.gov}
\emailAdd{mehtartani@bnl.gov}

\date{\today}
\abstract{The interplay between the small $x$ limit of QCD amplitudes and QCD factorization at moderate $x$ has been studied extensively in recent years. It was finally shown that semiclassical formulations of small $x$ physics can have the form of an infinite twist framework involving Transverse Momentum Dependent (TMD) distributions in the eikonal limit. In this work, we demonstrate that small $x$ distributions can be formulated in terms of transverse gauge links. This allows in particular for direct and efficient decompositions of observables into subamplitudes involving gauge invariant suboperators which span parton distributions.}


\date{\today}
\maketitle
\flushbottom

\section{Introduction\label{sec:intro}}

The two main regimes for a process with a hard scale $Q$ and center-of-mass
energy $s$ are the so-called Bjorken limit $Q\sim s \to \infty$ that relates to moderate values of $x$, where
QCD factorization applies, and the so-called Regge limit, for which $Q\ll s$ or $x\ll 1$. 
The main obstacle when studying the connections between QCD factorization
and QCD at small $x$ is the discrepancy in the involed non-perturbative
elements: QCD factorization involves parton distributions, whereas
the most general formulations of small $x$ physics involve the action
of Wilson line operators on hadronic states~\cite{McLerran:1993ni,McLerran:1993ka,McLerran:1994vd,Balitsky:1995ub,Balitsky:1998kc,Balitsky:1998ya}. In~\citep{Dominguez:2011wm}, the authors showed how to extract the small $x$ limit of
a Transverse Momentum Dependent (TMD) distribution from a Wilson line
operator. Their argument relies on two statements: the fact that the
longitudinal momentum fraction of partons in the distributions are
small in the eikonal limit, and the fact that transverse gluon fields
are subeikonal in the most commonly used gauges (see Section~\ref{sec:Yang-Mills}).

As an example, let us consider the Weizs\"{a}cker-Williams (WW) type gluon TMD operator for a gluon with longitudinal momentum fraction $x$ and transverse momentum $\k$ inside a hadron with a large momentum $P$ mostly in the $-$ ligthcone direction, which is defined as~\citep{Dominguez:2010xd}
\begin{equation}
\mathcal{O}^{ij}(x,\boldsymbol{k})=\int\frac{\rmd r^{+} \rmd^{2}\boldsymbol{r}}{(2\pi)^{3}}\rme^{ixP^{-}r^{+}-i(\boldsymbol{k}\cdot\boldsymbol{r})}F^{i-}(r)\,\mathcal{U}_{\left[r,0\right]}^{\left[+\right]}\,F^{j-}(0)\,\mathcal{U}_{[0,r]}^{\left[+\right]},\label{eq:dipdef}
\end{equation}
where $\mathcal{U}_{\left[r,0\right]}^{\left[\pm\right]}$ are staple-shaped
gauge links
\begin{equation}
\mathcal{U}_{\left[x,y\right]}^{\left[\pm\right]}= \left[x^+,\infty^+\right]_{\boldsymbol{x}} \left[\boldsymbol{x},\boldsymbol{y}\right]_{\pm\infty^+}\left[\pm\infty^+,y^{+}\right]_{\boldsymbol{y}}, \label{eq:stapledef}
\end{equation}
and $\left[x^+,\infty^+\right]_{\boldsymbol{x}}$ denotes a straight Wilson line in the fondamental representation of SU(3) along the $+$ direction at fixed transverse coordinate $\x$ and similarly for $\left[\pm\infty^+,y^{+}\right]_{\boldsymbol{y}}$. In particular, we have 
\beq\label{eq:L-wilson-line}
\left[x^+,y^+\right]_{\boldsymbol{x}} = U_\x(x^+,y^+)\equiv \cP \exp\left[ ig \int_{y^+}^{x^+} \rmd z^+ A^-(z^+,\x)\right]\,,
\eeq
where $A^-(x^+,\x)=n\cdot A $ is the $-$ component of the target gauge field evaluated on the light cone branch $x^-=0$, with $n=(1,0,0,1)$. Here it is understood that $A \equiv A^a t^a$  where $t^a$ are the QCD color matrices in the fundamental representation.

The transverse gauge link on the other hand is evaluated at constant light cone time $x^+$ 
\beq\label{eq:T-wilson-line}
\left[\x,\y\right]_{x^+} \equiv \cP_\lambda \exp\left[ -ig \int_{\y}^{\x} \rmd \z(\lambda) \cdot \A(x^+,\z(\lambda))\right]\,,
\eeq
where $\z(\lambda)\equiv (z^1,z^2)$ defines a trajectory in the transverse plane, that starts at $\y$ and ends at $\x$, and parametrized by the real number $0<\lambda<1$. 

Under the two hypothesis described above, one can neglect the phase
$ixP^{-}r^{+}$ in the Fourier transform in Eq.~(\ref{eq:dipdef}), as well as the transverse
part $\left[\boldsymbol{x},\boldsymbol{y}\right]_{\infty^+}$
of the staple-shaped gauge link. This observation allowed the authors
of~\citep{Dominguez:2010xd} to find a match between the WW-type TMD operator
and infinite Wilson line operators:
\begin{equation}
\mathcal{O}^{ij}(x\approx0,\boldsymbol{k})\propto\int \rmd^{2}\boldsymbol{r}\rme^{-i(\boldsymbol{k}\cdot\boldsymbol{r})}U_{\boldsymbol{r}}(\partial^{i}U^\dagger)_{\boldsymbol{r}}U_{\boldsymbol{0}}(\partial^{j}U^{\dagger})_{\boldsymbol{0}}.\label{eq:dipmatch}
\end{equation}
Such an operator appears naturally when one takes the first term in
a Taylor expansion of an observable at small $x$. The remarkable
equivalence in Eq.~(\ref{eq:dipmatch}) generated a lot of interest
for the physics of TMD distributions in the small $x$ community,
which gathered tremendous insight on these distributions from small
$x$ models~\cite{Metz:2011wb, Akcakaya:2012si, Dumitru:2016jku, Marquet:2016cgx, Boer:2017xpy, Marquet:2017xwy,Petreska:2018cbf, Altinoluk:2018byz}. Attempts have also been made in order to unify small $x$ and moderate $x$ evolution equations for TMDs~\citep{Balitsky:2015qba, Balitsky:2019ayf}. However, no equivalence was formed beyond the leading power in $\boldsymbol{k}/Q$ until very recently~\citep{Altinoluk:2019fui,Altinoluk:2019wyu}.

In~\citep{Altinoluk:2019fui,Altinoluk:2019wyu}, it was shown that a class of observables
at small $x$ could be entirely rewritten as the eikonal limit of
an infinite twist TMD framework. This new formulation of small-$x$ physics was based on a power expansion, then
the rearrangement of the expanded form by classifying terms depending
on the genuine twist of the non-perturbative operator involved, then
the resummation of power corrections to the accompanying Wilson coefficients.
Although the final expressions were fairly simple it would be cumbersome to generalize to other classes of observables.
It also relied on the assumption that neglecting the transverse gauge
links from the distribution in a gauge where transverse gluons are
subeikonal would not spoil QCD gauge invariance of the distribution.

In this article, we propose a more direct derivation of the equivalence
found in~\citep{Altinoluk:2019fui,Altinoluk:2019wyu} and uncover its underlying geometric structure which preserves the explicit gauge invariance of the operators. For this purpose,
we demonstrate that pairs of Wilson line operators have a powerful
formulation in terms of transverse gauge links built from rotated gluon
fields. In the new approach to TMD's at small $x$ in terms of transverse gauge link operators it will be straightforward  to generalize to other observables. 

The article is organized as follows: In Section~\ref{sec:Yang-Mills} we give a brief introduction of the semiclassical small $x$ basics and discuss gauge invariant for this framework. In Section~\ref{sec:par-transporter}, we show how pairs of Wilson lines can be interpreted as transverse gauge links thanks to parallel transports on the transverse plane and gauge invariance, and how this result has the form of a non-Abelian Stokes equation. In Section~\ref{sec:local}, we use the newly established form of the dipole operator to show how a power expansion can be systematically performed in a consistently gauge invariant way. As a second application of the results from Section~\ref{sec:par-transporter}, we show how to extract TMD suboperators from the dipole in Section~\ref{sec:DIS}, and in Section~\ref{sec:gen} we extend this method to more generic 2-Wilson-line operators. Finally, we extend the method to a 3-line operator in Section~\ref{sec:3L}.

\section{Background field at high energy\label{sec:Yang-Mills}}

Consider a hadronic target moving in the negative $z$ direction,
close to the light cone, i.e.,  $x^-\sim 0$. It can be described by a classical current~\citep{McLerran:1994vd,Gelis:2010nm}
\begin{equation}
J^{-}(x)\approx J^{-}(x^{+},\boldsymbol{x}),\quad\text{and}\quad J^{+}\approx J^{i}\approx0\,,\label{eq:current}
\end{equation}
that generates a gauge field which only depends on light cone time
$x^{+}$ and the transverse coordinate $\boldsymbol{x}$. In such
a framework, it turns out that both covariant $\partial\cdot A=0$
and light cone $A^{+}=0$ gauge share a common solution. Indeed, it
immediately follows from $A^{+}=A^{i}=0$ and the independence on
$x^{-}$ that $\partial\cdot A=\partial^{+}A^{-}=0.$ The equation
of motion for the field reads
\begin{equation}
\left[D_{\mu},F^{\mu-}\right]=-\partial^{i}F^{i-}=-\boldsymbol{\partial}^{2}A^{-}=J^{-}.\label{eq:YM}
\end{equation}
where 
\beq 
F^{\mu\nu} \equiv \del^\mu A^\nu-\del^\nu A^\mu-ig[A^\mu,A^\nu]\quad \text{and} \quad D^\mu \equiv \del^\mu-ig A^\mu\,,
\eeq
denote the field strength tensor and the covariant derivative. With the above choice of gauges the current is covariantly conserved since  $D^+J^-=\partial^{+}J^{-}=0$. Furthermore,  note that although
$A^{-}$ obeys a Poisson equation, it is an exact solution of the
Yang-Mills equations. 

In the following, we will make use of the existence
of a residal gauge freedom in $A^{+}=0$ light cone gauge: all gauge
transformations with the form
\begin{align}
A^{-}(x^{+},\boldsymbol{x}) & \rightarrow\Omega_{\boldsymbol{x}}(x^{+})A^{-}(x^{+},\boldsymbol{x})\Omega_{\boldsymbol{x}}^{-1}(x^{+})-\frac{1}{ig}\Omega_{\boldsymbol{x}}(x^{+})\partial^{-}\Omega_{\boldsymbol{x}}^{-1}(x^{+})\nonumber \\
A^{i}(x^{+},\boldsymbol{x}) & \rightarrow-\frac{1}{ig}\Omega_{\boldsymbol{x}}(x^{+})\partial^{i}\Omega_{\boldsymbol{x}}^{-1}(x^{+}),\label{eq:gt-cov}
\end{align}
where $\Omega_\x(x^{+})$ is an element of the gauge group
SU(3) that preserves the condition $A^{+}=0$. Because $A^+$ is suppressed on the branch $x^-=0$ the above decomposition is expected to span a larger subset of SU(3) beyond light cone gauge. 

Under such a gauge transformation,
the Wilson line
\begin{equation}
U_{\boldsymbol{x}}(\xi_{2},\xi_{1})=\mathcal{P}\exp\left[ig\int_{\xi_{1}}^{\xi_{2}}\!\rmd x^+A^{-}(x^+,\boldsymbol{x})\right]\label{eq:wilson-line}
\end{equation}
transforms as
\begin{equation}
U_{\boldsymbol{x}}(\xi_{2},\xi_{1})\,\rightarrow\,\Omega_{\boldsymbol{x}}(\xi_{2})U_{\boldsymbol{x}}(\xi_{2},\xi_{1})\Omega_{\boldsymbol{x}}^{-1}(\xi_{1}).\label{eq:wl-trans}
\end{equation}
It is possible to choose the transformation in order to cancel $A^{-}$ altogether,
with
\begin{equation}
\Omega_{\boldsymbol{x}}(\xi)=\Omega_{\boldsymbol{x}}(\infty)U_{\boldsymbol{x}}(\infty,\xi),\quad\text{and}\quad\Omega_{\boldsymbol{x}}^{-1}(\xi)=U_{\boldsymbol{x}}(\xi,-\infty)\Omega_{\boldsymbol{x}}^{-1}(-\infty).\label{eq:gt-cov-null}
\end{equation}
Note that there is an infinite number of gauge transformations, spanning
the entire gauge groupe SU(3), that satisfy~(\ref{eq:gt-cov-null}).
This freedom in fixing the boundary conditions translates into the
different regularization schemes for the spurious $1/k^{+}$ singularity
encountered in light cone gauge, see for example~\citep{Chirilli:2015fza}.

\section{Transverse QCD strings and the dipole operator\label{sec:par-transporter}}

In the standard semi-classical formulations of small $x$ physics, scattering amplitudes are most commonly computed in $A^+=0$ light cone gauge by fixing the residual gauge freedom such that the classical transverse field vanishes. Hence, only the $-$ component of the background field is taken into account through lightlike Wilson line operators along the $x^+$ direction. \\
The connection to TMD physics requires the formulation of the problem in terms of field strength tensors, typically introduced by performing a gradient expansion in transverse coordinate space that yields $\partial^i A^- \sim F^{i-}$. In effect, this corresponds to the parallel transport of Wilson line operators on the transverse plane as will be shown shortly. \\
\subsection{Parallel transport on the transverse plane}
Let us first show how gauge rotations can be rewritten as transverse
gauge links. For any pair of transverse positions (\textbf{$\boldsymbol{x}_{1},\boldsymbol{x}_{2}$}) and
defining $\boldsymbol{r}=\boldsymbol{x}_{1}-\boldsymbol{x}_{2}$, we can readily write\footnote{From now on, we will denote light cone times with standard time notations $\xi$ or $t$ for reader's convenience.}
\beq
\Omega_{\boldsymbol{x}_{1}}(t)&=&\Omega_{\boldsymbol{x}_{2}}(t)+\int_{0}^{1}\rmd s \, \frac{\rmd }{\rmd s}\Omega_{\boldsymbol{x}_{2}+s\boldsymbol{r}}(t)\nn
&=&\Omega_{\boldsymbol{x}_{2}}(t)-\r^{i}\int_{0}^{1}\rmd s \, (\partial^{i}\Omega)_{\boldsymbol{x}_{2}+s\boldsymbol{r}}(t)\,.\label{eq:par-transport}
\eeq
Now, note that
\begin{equation}
\partial^{i}\Omega_{\boldsymbol{x}}(t)=ig {A}^{i}(t,\boldsymbol{x})\Omega_{\boldsymbol{x}}(t),\label{eq:igA}
\end{equation}
where ${A}^{i}$ (with $i=1,2$)  is the pure gauge field obtained from the rotation $\Omega$ (see Eq.~(\ref{eq:gt-cov})). Combining
the two remarks above then multiplying Eq.~(\ref{eq:par-transport}) by $\Omega_{\boldsymbol{x}_{2}}^{-1}(t)$
on the right yields:
\begin{equation}
\Omega_{\boldsymbol{x}_{1}}(t)\Omega_{\boldsymbol{x}_{2}}^{-1}(t)=1-ig \, r^{i}\int_{0}^{1} \rmd s\, A^i(t,\boldsymbol{x}_{2}+s\boldsymbol{r})\, \Omega_{\boldsymbol{x}_{2}+s\boldsymbol{r}}(t)\Omega_{\boldsymbol{x}_{2}}^{-1}(t).\label{eq:par-transport-1}
\end{equation}
The final step is now to notice that Eq.~(\ref{eq:par-transport-1})
is the equation which defines a Wilson line along the straight line trajectory parametrized by the real number $s$ with values between 0 and 1 and such that $\boldsymbol{z}(0)=\boldsymbol{x}_2$ and $\boldsymbol{z}(1)=\boldsymbol{x}_1$, that is:
$$\boldsymbol{z}(s)=\boldsymbol{x}_{2}+s\boldsymbol{r}$$
This Wilson line, that we denote as $[\boldsymbol{x}_1,\boldsymbol{x}_2]_t$, solves the following equation:
\begin{equation}
[\boldsymbol{x}_{1},\boldsymbol{x}_{2}]_{t}=1-ig\!\int_{\boldsymbol{x}_{2}}^{\boldsymbol{x}_{1}}\!\!\rmd \z\cdot\A(t,\boldsymbol{z})[\boldsymbol{z},\boldsymbol{x}_{2}]_{t}\,.\label{eq:linkdefL}
\end{equation}
Although the transverse gauge link was constructed for a straight line trajectory it can be easily shown that it is independent of the trajectory connecting the endpoints so long as the transverse field is a pure gauge. \\
By comparing \eqn{eq:par-transport-1} and \eqn{eq:linkdefL} we can make the following identification:
\begin{equation}
\Omega_{\boldsymbol{x}_{1}}(t)\Omega_{\boldsymbol{x}_{2}}^{-1}(t)=[\boldsymbol{x}_{1},\boldsymbol{x}_{2}]_{t}.\label{eq:translink}
\end{equation}
An equivalent relation can readily be obtained:
\begin{equation}
[\boldsymbol{x}_{1},\boldsymbol{x}_{2}]_{t}=1-ig\!\int_{\boldsymbol{x}_{2}}^{\boldsymbol{x}_{1}}[\boldsymbol{x}_{1},\boldsymbol{z}]_{t}\, \boldsymbol{A}(t,\boldsymbol{z})\cdot\rmd\z .\label{eq:linkdefR}
\end{equation}

\subsection{Dipole operator and the non-Abelian Stokes theorem}

\begin{figure}[H]
\begin{center}
\includegraphics[width=0.9\linewidth]{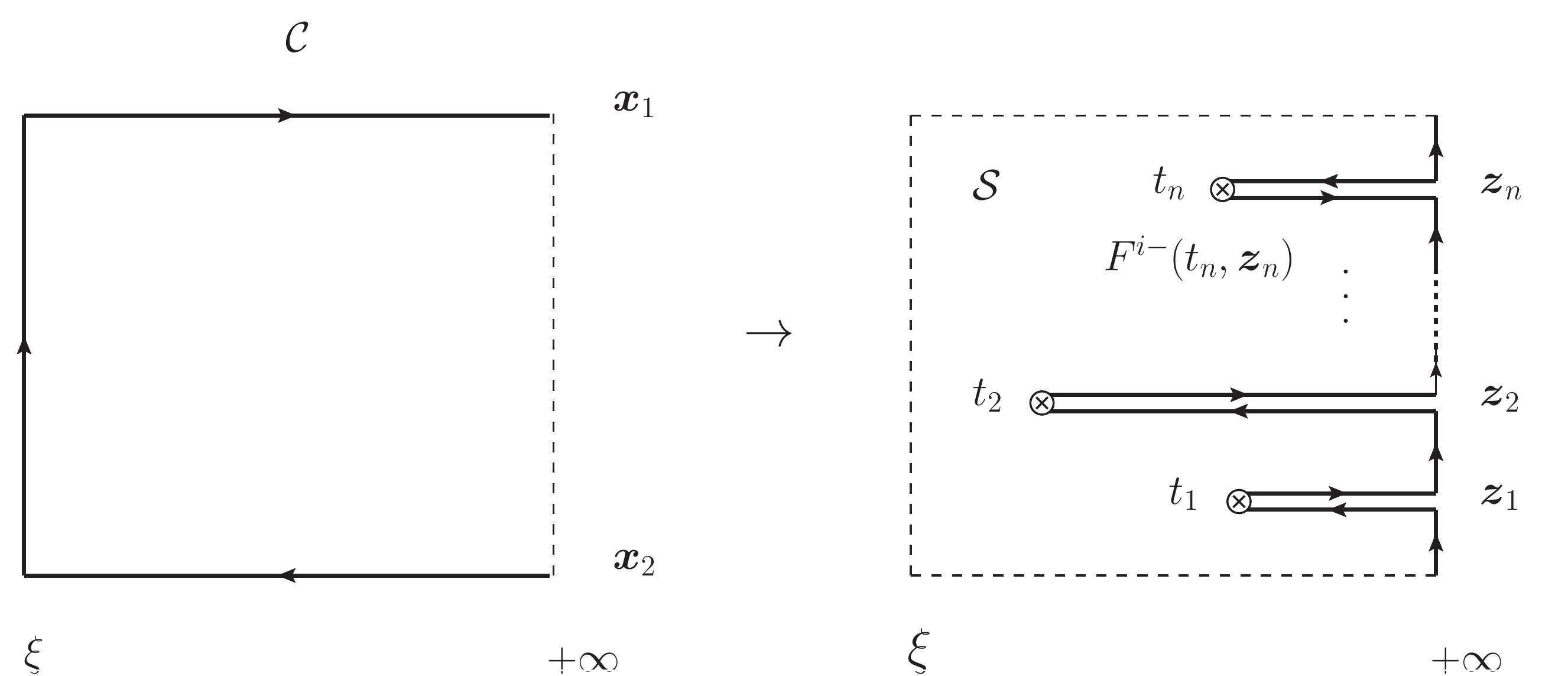}
\caption{Non-Abelian Stokes theorem \label{fig:stokes}}
\end{center}
\end{figure}

Now let us consider the (non-singlet) dipole operator, with gauge
links at infinity $\Omega_{\boldsymbol{x}_{1,2}}(\infty)$ fully accounted
for:
\begin{equation}
\left(\mathcal{O}_{\xi}(\boldsymbol{x}_{1},\boldsymbol{x}_{2})\right)_{ij}\equiv\left(\Omega_{\boldsymbol{x}_{1}}(\infty)U_{\boldsymbol{x}_{1}}(\infty,\xi)U_{\boldsymbol{x}_{2}}^{\dagger}(\xi,\infty)\Omega_{\boldsymbol{x}_{2}}^{-1}(\infty)\right)_{ij}\,.\label{eq:dipole}
\end{equation}
In most cases at small $x$ it is assumed that the classical field
$A^{-}$ has a compact support which is very peaked around $x^+=0$ in \eqn{eq:wilson-line}. As a result, so long as $x^+<0$ one can replace $\xi$ by $-\infty$ in our expressions and one can deal with infinite Wilson lines. Here, we will keep a generic $\xi$. Absorbing the gauge
links at infinity $\Omega_{\boldsymbol{x}_{1}}(\infty)$ and $\Omega_{\boldsymbol{x}_{2}}^{-1}(\infty)$
into the Wilson lines with the help of relations like Eq.~(\ref{eq:gt-cov}), the dipole operator becomes
\begin{equation}
\mathcal{O}_{\xi}(\boldsymbol{x}_{1},\boldsymbol{x}_{2})\rightarrow U_{\boldsymbol{x}_{1}}(\infty,\xi)\Omega_{\boldsymbol{x}_{1}}(\xi)\Omega_{\boldsymbol{x}_{2}}^{-1}(\xi)U_{\boldsymbol{x}_{2}}^{\dagger}(\xi,\infty),\label{eq:dipolerot}
\end{equation}
where the lines are now built from gauge-rotated gluon fields from
Eq.~(\ref{eq:gt-cov}). Using Eq.~(\ref{eq:translink})
finally allows to write the dipole operator with transverse gauge
links:
\begin{align}
\mathcal{O}_{\xi}(\boldsymbol{x}_{1},\boldsymbol{x}_{2}) & =U_{\boldsymbol{x}_{1}}(\infty,\xi)[\boldsymbol{x}_{1},\boldsymbol{x}_{2}]_{\xi}U_{\boldsymbol{x}_{2}}^{\dagger}(\xi,\infty).\label{eq:dipoleinv}
\end{align}
A diagrammatic depiction of $\mathcal{O}_{\xi}(\boldsymbol{x}_{1},\boldsymbol{x}_{2})$  is given in Fig.~\ref{fig:stokes}, left panel. 
In Eq.~(\ref{eq:dipoleinv}), the transverse gauge link is built
from the pure gauge transverse gluons $A^{i}(x^{+},\boldsymbol{x})=-\frac{1}{ig}\Omega_{\boldsymbol{x}}(x^{+})\partial^{i}\Omega_{\boldsymbol{x}}^{-1}(x^{+}).$
It is possible to absorb the longitudinal Wilson lines into the transverse link with the
following change of variables
\begin{equation}
{A}^{i}(\xi,\boldsymbol{z})\rightarrow\hat{A}^{i}(\xi,\boldsymbol{z})\equiv U_{\boldsymbol{z}}(\xi,\infty){A}^{i}(\xi,\boldsymbol{z})U_{\boldsymbol{z}}^{\dagger}(\xi,\infty)+\frac{1}{ig}(\partial^{i}U_{\boldsymbol{z}})(\infty,\xi)U^\dag_{\boldsymbol{z}}(\xi,\infty).\label{eq:Airot}
\end{equation}
Note that the hatted field is non-local in $x^+$ and transforms similarly to the field strength tensor under a gauge rotation, i.e.
\beq
 \hatA^{i}(\xi,\z) \to  \Omega_\z(\infty)\hatA^{i} (\xi,\z) \Omega_\z^{-1}(\infty).  
\eeq
We can actually trade the dependence on the gauge field with that
of the field strength tensor with simple algebra:
\begin{equation}
\hat A^{i}(\xi,\boldsymbol{z})=\int_{\xi}^{\infty}\!\!\rmd t\,U_{\boldsymbol{z}}(\infty,t)F^{i-}(t,\boldsymbol{z})U_{\boldsymbol{z}}^{\dagger}(t,\infty)+ A^i(\infty,\boldsymbol{z}).\label{eq:AiF}
\end{equation}
The boundary term $A^i(\infty,\boldsymbol{z})$ can be absorbed into a transverse gauge link at $t=\infty$. \\
Making use of the Fierz Identity 
\beq\label{eq:Fierz}
U{\bt}^a U^\dag =  {\bt}^b W^{ba} \,,
\eeq
one can express \eqn{eq:AiF} in terms of a single Wilson line $W^{ba}$ in the adjoint representation 
\begin{equation}
\hat A^{i}(\xi,\boldsymbol{z})=\int_{\xi}^{\infty}\!\!\rmd t\,{\bt}^b\,W^{ba}_{\boldsymbol{z}}(\infty,t)F^{a,i-}(t,\boldsymbol{z}) + A^i(\infty,\boldsymbol{z}).\label{eq:AiF}
\end{equation}
For the sake of clarity, we will distinguish the gauge links which
depend on the rotated field from Eqs.~(\ref{eq:Airot}) and  (\ref{eq:AiF})
from the regular gauge links by hatting their coordinates. We have
now established that
\begin{align}
\mathcal{O}_{\xi}(\boldsymbol{x}_{1},\boldsymbol{x}_{2}) & = [\hat{\boldsymbol{x}}_{1},\hat{\boldsymbol{x}}_{2}]_{\xi},\label{eq:dipoleTrans}
\end{align}
which allows to understand the dipole operator as a transverse string
built from the gluons in Eq.~(\ref{eq:AiF}).  \\
We can rewrite this result in the form of the non-Abelian Stokes theorem~(\cite{Fishbane:1980eq}, see also~\cite{Wiedemann:2000ez}): the integral of $A^\mu$ over the square contour $\mathcal{C}$ on the left hand side of Fig.~\ref{fig:stokes} is equal to the integral of the so-called \textit{twisted} strength tensor $U F^{\mu\nu}U^\dagger$ inside the surface $\mathcal{S}$ defined by this contour:
$$ \mathcal{S} = \{(t,\boldsymbol{z}) ; t \in [\xi,\infty],\boldsymbol{z}\in [\boldsymbol{x}_1,\boldsymbol{x}_2]\}. $$ \\
Indeed, the more explicit Eq.~(\ref{eq:dipoleTrans}), with non-vanishing transverse fields at infinite light cone time, is given by
\begin{align}
 & \mathcal{P} \rme^{ig\int_{\xi}^{\infty}\rmd tA^{-}(t,\boldsymbol{x}_{1})}\mathcal{P}^{-ig\int_{\boldsymbol{x}_{2}}^{\boldsymbol{x}_{1}}\rmd \boldsymbol{z}\cdot \boldsymbol{A}(\xi,\boldsymbol{z})}\mathcal{P}\rme^{ig\int_{\xi}^{\infty}\rmd tA^{-}(t,\boldsymbol{x}_{2})}\mathcal{P}\rme^{-ig\int_{\boldsymbol{x}_{1}}^{\boldsymbol{x}_{2}}\rmd \boldsymbol{z}\cdot \boldsymbol{A}(\infty,\boldsymbol{z})}\label{eq:stokes}\\
 & =\mathcal{P}\exp\left[-ig\int_{\boldsymbol{x}_{2}}^{\boldsymbol{x}_{1}}\rmd \boldsymbol{z}^i\int_{\xi}^{\infty} \! \rmd t \, [\x_1,\z]_\infty\,[\infty,t]_{\boldsymbol{z}}F^{i-}(t,\boldsymbol{z})[t,\infty]_{\boldsymbol{z}}[\z,\x_1]_\infty\right], \nonumber 
\end{align}
which is a form of the Stokes equation
\begin{equation}
\mathcal{P}\exp\left[\oint_{\mathcal{C}}\rmd x_\mu A^\mu(x)\right]= \mathcal{P} \exp\left[\int_{\mathcal{S}} \rmd \sigma_{\mu\nu} \, UF^{\mu\nu}U^\dagger \right],
\end{equation}
where $\rmd\sigma_{\mu\nu}$ is the surface measure on $\mathcal{S}$ and $U$ denotes a Wilson line connecting the point $x \in {\cal S}$ enclosed by the surface measure to an arbitrary base point $O$ on $\cal C$, which in Eq.~(\ref{eq:stokes}) is taken to be $O\equiv (\infty,\x_1)$.

\section{Gauge invariant power expansion: local formulation \label{sec:local}}
In this section we shall use the formulation of the dipole operator established above in order to construct a power expansion of the dipole operator \eqn{eq:dipoleTrans} that is explicitly gauge invariant order by order in powers of $\r=\x_2-\x_1$. 

First, let us show the following expression for the non-singlet dipole~(\ref{eq:dipole}):
\beq
\left(\cO(\b,\b-\r)\right)_{ij} = \left(\rme^{-\r\cdot (\bdel_b -i g \hatA(\b))}\right)_{ij}. \label{Or}
\eeq
where $\hatA(\b)$ was previously introduced in Eq.~(\ref{eq:Airot}). Similarly, one can show that
\beq
\cO(\b+\r,\b) = \rme^{\r\cdot (\overleftarrow{\bdel}_b -i g \hatA(\b))}.\label{Ol}
\eeq

Our starting point is \eqn{eq:dipole} expressed in the variables $\r$ and $\b$ with $\xi=-\infty$ (the generalization to arbitrary initial points is straightforward)
\beq 
\mathcal{O}(\b,\b-\r)&=&  \left(U(\infty,-\infty) \Omega(-\infty) \right)_{\b}  \left( \Omega^{-1}(-\infty)U^{\dagger}(-\infty,\infty)\right)_{\b-\r}\nn
&=&  \left(U(\infty,-\infty) \Omega(-\infty)\right)_{\b} \, \rme^{-\r\cdot\bdel_b} \,  \left(\Omega^{-1}(-\infty) U^{\dagger}(-\infty,\infty)\right)_{\b}\label{eq:dipole}
\eeq
where in the last line we have introduced the translation operator $\rme^{-\r\cdot\bdel_b}$. The transverse derivative can be turned into a covariant one by absorbing the gauge link into the translation operator as follows (omitting the light-cone time dependence to alleviate the notations, i.e., $\Omega \equiv \Omega(-\infty) $ and $U \equiv U(\infty,-\infty)$),
\beq
\bdel_b  \Omega^{-1}  U^{\dagger}  \phi(\b) &=&\Omega^{-1}  U^{\dagger}  \left[ \del_b  +U \Omega  \left(\del_b  \Omega^{-1}  U^{\dagger}\right) \right]  \phi(\b) \nn 
&=&\Omega^{-1}  U^{\dagger}  \left[ \bdel_b   + U \Omega    \left(\bdel_b  \Omega^{-1} \right)  U^{\dagger} +U  \left(\bdel_b  U^{\dagger} \right) \right]  \phi(\b)\nn
&=&\Omega^{-1}  U^{\dagger}  \left( \bdel_b   - ig \hatA(\b) \right) \phi(\b)  
\eeq
where $\phi(\b)$ is a test function.  By applying the above commutation relation it follows that 
\beq
 U \Omega  \, ( \bdel_b  )^n\Omega^{-1}  U^{\dagger}  \phi(\b)=  \left( \bdel_b   - ig \hatA(\b) \right)^n \phi(\b).
\eeq
This amounts to replacing the derivative over $\b$ in the translation operator by the hatted transverse covariant derivative 
\beq
\hat \D _b  \equiv \bdel_b - ig \hatA(\b),
\eeq
that involves both $A^-$ and $A^i$ fields, and leads to Eqs~(\ref{Or}) and (\ref{Ol}). These two equations can be used in order to perform a local expansion, thus an expansion in powers of small transverse momenta, with explicit gauge invariance of all operators at each step of the process. For this purpose, one can use the following identity which is a direct consequence of Eq~(\ref{eq:Airot})\footnote{Here the covariant derivative $D^i_\b U_\b$ is to be understood as an operator acting on everything on its right, not to be mistaken for $(D^i U)_\b$.}
\beq
 \del^i_\b-\hat A^i(t_0,\b) = U_\b(+\infty,t_0) \,D^i(t_0,\b)\,  U_\b^\dag(t_0,+\infty).\label{eq:DieqUDU}
\eeq
Let us detail the steps for the second term of the local expansion of Eq~(\ref{Or}). First, use the definition of the hatted field on the right:
\begin{align}
 & \mathcal{O}^{\left(2\right)}\left(\boldsymbol{b},\boldsymbol{b}-\boldsymbol{r}\right)\label{eq:O2}\\
 & =\boldsymbol{r}^{i_{1}}\boldsymbol{r}^{i_{0}}\left(\partial_{\boldsymbol{b}}-ig\hatA\left(-\infty,\boldsymbol{b}\right)\right)^{i_{1}}\left(\partial_{\boldsymbol{b}}-ig\hatA\left(-\infty,\boldsymbol{b}\right)\right)^{i_{0}}\nonumber \\
 & =-ig\int_{-\infty}^{+\infty}\rmd t_{0}\, \boldsymbol{r}^{i_{1}}\boldsymbol{r}^{i_{0}}\left(\partial_{\boldsymbol{b}}-ig\hatA\left(-\infty,\boldsymbol{b}\right)\right)^{i_{1}}\left[U_{\boldsymbol{b}}\left(\infty,t_{0}\right)F^{i_0-}\left(t_{0},\boldsymbol{b}\right)U_{\boldsymbol{b}}^{\dagger}\left(t_{0},\infty\right)\right]\nonumber 
\end{align}
The trick is now to write the hatted field on the left, which is evaluated at $-\infty$ light cone time, as a function of the hatted field at light cone time $t_0$. Thus we need to use the following relation:
\beq\label{eq:hatA-F-1}
 \hat A^i(-\infty,\b) \equiv   \hat A^i(t_0,\b) + \int_{-\infty}^{t_0}\rmd t \, U(+\infty,t)\,F^{i-}(t,\b) \, U^\dag(t,+\infty),
\eeq
which, when combined with Eq~(\ref{eq:DieqUDU}), leads to
\begin{align}
 & \left(\partial_{\boldsymbol{b}}-ig\hatA\left(-\infty,\boldsymbol{b}\right)\right)^{i_{1}}\label{eq:Di1}\\
 & =\partial_{\boldsymbol{b}}^{i_{1}}-ig\hatA^{i_{1}}\left(t_{0},\boldsymbol{b}\right)-ig\int_{-\infty}^{t_{0}}\rmd t_{1}U_{\boldsymbol{b}}\left(\infty,t_{1}\right)F^{i_{1}-}\left(t_{1},\boldsymbol{b}\right)U_{\boldsymbol{b}}^{\dagger}\left(t_{1},\infty\right)\nonumber \\
 & =U_{\boldsymbol{b}}\left(\infty,t_{0}\right)D^{i_{1}}\left(t_{0},\boldsymbol{b}\right)U_{\boldsymbol{b}}^{\dagger}\left(t_{0},\infty\right)-ig\int_{-\infty}^{t_{0}}\!\rmd t_{1}U_{\boldsymbol{b}}\left(\infty,t_{1}\right)F^{i_{1}-}\left(t_{1},\boldsymbol{b}\right)U_{\boldsymbol{b}}^{\dagger}\left(t_{1},\infty\right)\nonumber 
\end{align}
We can then conclude with
\begin{align}
 & \mathcal{O}_2 \label{eq:O2f} =-ig\,  \boldsymbol{r}^{i_{1}}\, \boldsymbol{r}^{i_{0}}\int_{-\infty}^{+\infty}\rmd t_{0}\left[\infty,t_{0}\right]_{\boldsymbol{b}}D^{i_{1}}\left(t_{0},\boldsymbol{b}\right)\left(F^{i_{0}-}\left(t_{0},\boldsymbol{b}\right)\left[t_{0},\infty\right]_{\boldsymbol{b}}\right) \\
 & -g^{2}\, \boldsymbol{r}^{i_{1}}\, \boldsymbol{r}^{i_{0}}\int_{-\infty}^{+\infty}\rmd t_{0}\int_{-\infty}^{t_{0}}\rmd t_{1}\left[\infty,t_{1}\right]_{\boldsymbol{b}}F^{i_{1}-}\left(t_{1},\boldsymbol{b}\right)\left[t_{1},t_{0}\right]_{\boldsymbol{b}}F^{i_{0}-}\left(t_{0},\boldsymbol{b}\right)\left[t_{0},\infty\right]_{\boldsymbol{b}}.\nonumber 
\end{align}
By recursion with similar steps, one can prove the general form of $|\boldsymbol{r}|^n$ term:
\beq
{ \cal O}_n &=&  (-1)^n \boldsymbol{r}^{i_m} ...\boldsymbol{r}^{i_1}  \boldsymbol{r}^{i_0}  \sum_{k_0 ...k_m}\int_{-\infty}^{\infty} \rmd t_0\int_{-\infty}^{t_0}\rmd t_1\,...\int_{-\infty}^{t_{m-1}}\rmd t_m \\
&\times& \,U(+\infty,t_m)\,(\r \cdot D)^{k_m} F^{i_m-}(t_m)U^\dag(t_m,t_{m-1}) \,... \, U^\dag(t_1,t_0) \, (\r \cdot D)^{k_0}\, F^{i_0-}(t_0) U^\dag(t_0,+\infty) \nonumber
\eeq
where the sum over $k_0 ...k_m$ is constrained by
\beq 
\sum_{j=0} k_j =n-m
\eeq
In simple words, the $n$-th term in the gauge invariant local expansion of the dipole operator is the sum of all possible insertions of $F$ tensors and covariant derivatives with the appropriate gauge links, such that the number of $F$'s and the number of $D$'s sum up to $n$.
As an illustration, the first few orders read: 
\beq
 { \cal O}^{i}_1=  - \r^i \int_{-\infty}^{+\infty} \rmd t \,[+\infty,t] F^{i-}(t) [t,+\infty]\,,
\eeq
for the first order 
\beq
 { \cal O}^{ij}_2 &=&  \r^i \r^j \int_{-\infty}^{+\infty} \rmd t_1\,\int_{-\infty}^{t_1} \rmd t_2 \, [+\infty,t_2] F^{i-} [t_2,t_1] F^{j-} [t_1,+\infty] \nn
 &+&  \, \int_{-\infty}^{+\infty} \rmd t \, [+\infty,t] \, D^j F^{i-}(t) \, [t,+\infty]\,, 
\eeq
for the second, and for the third we have
\beq
 { \cal O}^{ijk}_3 &=&  -\r^i \r^j \r^k \int_{-\infty}^{+\infty} \rmd t_1\,\int_{-\infty}^{t_1} \rmd t_2 \,  \,\int_{-\infty}^{t_2} \rmd t_3 \,  [\infty,t_3] F^{i-} [t_3,t_2]F^{j-} [t_3,t_2] F^{k-} [t_1,+\infty] \nn
 &+&  \, \int_{-\infty}^{+\infty} \rmd t_1\,\int_{-\infty}^{t_1} \rmd t_2 \, [+\infty,t_2] \left(  D^k F^{i-} [t_2,t_1] F^{j-}   + F^{i-} [t_2,t_1] D^k  F^{j-} \right)[t_1,+\infty] \nn
 &+&  \, \int_{-\infty}^{\infty} \rmd t \,[+\infty,t] D^j D^k F^{i-}(t) \, [t,+\infty]\,.
\eeq
The formulation of the dipole operator as a transverse QCD string can thus be used in order to perform the power expansion of a small $x$ observable while keeping explicitely gauge invariant operators. This is a particularly difficult task when using more standard forms of the small $x$ observables, where the gluon field strength tensor only appears via the derivative of Wilson lines thanks to the relation $\partial^i A^- = F^{i-}$. For example, see~\cite{Dumitru:2016jku} where gauge invariance, while not broken, is not explicitely respected due to the presence of double derivatives of Wilson lines which lead to simple derivatives of fields. Here, we established a systematic framework to perform such expansions with explicit invariance. In the following section, we will reproduce and generalize the results of~\cite{Altinoluk:2019wyu} which allow for another gauge invariant power expansion where the gauge invariant bilocal operators are left untouched while hard subamplitudes undergo a Taylor expansion instead.

\section{Application to the small-$x$/TMD equivalence for DIS dijet production\label{sec:DIS}}

\begin{figure}[H]
\begin{center}
\includegraphics[width=.5\linewidth]{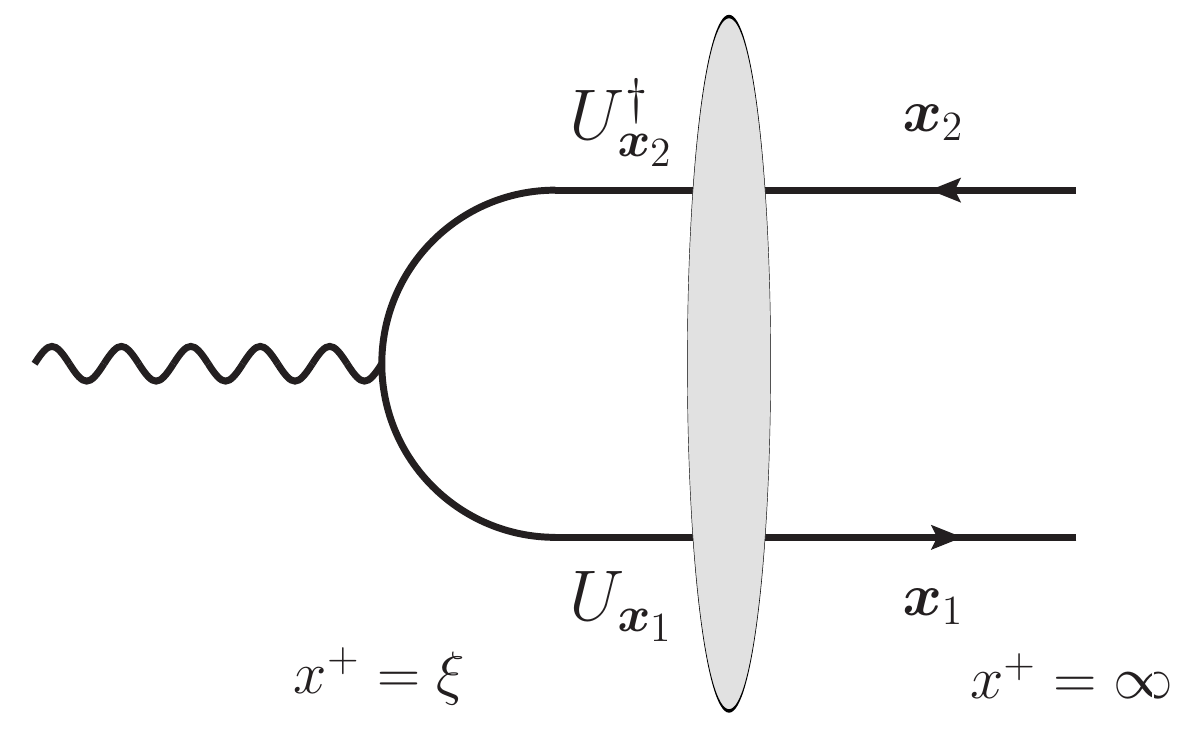}
\end{center}
\caption{The amplitude for DIS dijet production involves the non-singlet dipole from Eq.~(\ref{eq:dipole}). Gray blobs represent the interactions with the classical external field, which effectively dresses the quark and the antiquark with Wilson lines to build the dipole operator.}
\end{figure}

Starting from Eq.~(\ref{eq:dipoleTrans}), one can apply successively
both relations (\ref{eq:linkdefL}) and (\ref{eq:linkdefR}) in order
to rewrite the dipole operator into 1-body and 2-body contributions:
\begin{align}
\mathcal{O}_{\xi}(\boldsymbol{x}_{1},\boldsymbol{x}_{2}) & =1-ig\!\int_{\boldsymbol{x}_{2}}^{\boldsymbol{x}_{1}}\!\!\rmd \boldsymbol{z}^i\hatA^i(\xi,\boldsymbol{z})\label{eq:trick}\\
 & +(ig)^{2}\!\int_{\boldsymbol{x}_{2}}^{\boldsymbol{x}_{1}}\!\!\rmd \boldsymbol{z}^i \!\int_{\boldsymbol{x}_{2}}^{\boldsymbol{z}}\!\!\rmd \boldsymbol{z}^{\prime j}\hatA^i(\xi,\boldsymbol{z})[\hat{\boldsymbol{z}},\hat{\boldsymbol{z}}^{\prime}]_{\xi}\hatA^j (\xi,\boldsymbol{z}^{\prime})\nonumber 
\end{align}
 Then, recalling the explicit expression for the rotated fields in terms of twisted stength tensors:
\begin{align}
\mathcal{O}_{\xi}(\boldsymbol{x}_{1},\boldsymbol{x}_{2}) & =1+ig\!\int_{\xi}^{\infty}\!\!\rmd t\int_{\boldsymbol{x}_{2}}^{\boldsymbol{x}_{1}}\!\!\rmd \boldsymbol{z}^iU_{\boldsymbol{z}}(\infty,t)F^{-i}(t,\boldsymbol{z})U_{\boldsymbol{z}}^{\dagger}(\infty,t)\label{eq:trick-1}\\
 & +(ig)^{2}\!\int_{\xi}^{\infty}\!\!\rmd t\int_{\xi}^{\infty}\!\!\rmd t^{\prime}\int_{\boldsymbol{x}_{2}}^{\boldsymbol{x}_{1}}\!\!\rmd \boldsymbol{z}^i\!\int_{\boldsymbol{x}_{2}}^{\boldsymbol{z}}\!\!\rmd \boldsymbol{z}^{\prime j} \nonumber \\
 & \times U_{\boldsymbol{z}}(\infty,t)F^{-i}(t,\boldsymbol{z})U_{\boldsymbol{z}}^{\dagger}(\infty,t)[\hat{\boldsymbol{z}},\hat{\boldsymbol{z}}^{\prime}]_{t}U_{\boldsymbol{z}^{\prime}}(\infty,t^{\prime})F^{-j}(t^{\prime},\boldsymbol{z}^{\prime})U_{\boldsymbol{z}^{\prime}}^{\dagger}(\infty,t^{\prime}).\nonumber 
\end{align}
Using the expression for the hatted links
\begin{equation}
[\hat{\boldsymbol{z}},\hat{\boldsymbol{z}}^{\prime}]_{\xi}=U_{\boldsymbol{z}}(\infty,\xi)[\boldsymbol{z},\boldsymbol{z}^{\prime}]_{\xi}U_{\boldsymbol{z}^{\prime}}^{\dagger}(\infty,\xi),\label{eq:hattedlink}
\end{equation}
as well as 
\begin{equation}
U_{\boldsymbol{z}}^{\dagger}(\infty,t)U_{\boldsymbol{z}}(\infty,\xi)=[t,\xi]_{\boldsymbol{z}},\label{eq:translongi}
\end{equation}
and its counterpart for $\left[\xi,t^{\prime}\right]_{\boldsymbol{z}^{\prime}}$,
the dipole operators ends up entirely rewritten as the sum of 1-body
and 2-body operators, in an explicitely gauge invariant way:
\begin{align}
\mathcal{O}_{\xi}(\boldsymbol{x}_{1},\boldsymbol{x}_{2}) & =1+ig\!\int_{\xi}^{\infty}\!\!\rmd t\int_{\boldsymbol{x}_{2}}^{\boldsymbol{x}_{1}}\!\!\rmd \boldsymbol{z}^i \,[\infty,t]_{\boldsymbol{z}}F^{-i}(t,\boldsymbol{z})[t,\infty]_{\boldsymbol{z}}\label{eq:gaugeinvO}\\
 & +(ig)^{2}\!\int_{\xi}^{\infty}\!\!\rmd t\int_{\xi}^{\infty}\!\!\rmd t^{\prime}\int_{\boldsymbol{x}_{2}}^{\boldsymbol{x}_{1}}\!\!\rmd \boldsymbol{z}^i\!\int_{\boldsymbol{x}_{2}}^{\boldsymbol{z}}\!\!\rmd \boldsymbol{z}^{\prime j}\nonumber \\
 & \times [\infty,t]_{\boldsymbol{z}}F^{-\alpha}(t,\boldsymbol{z})[t,\xi]_{\boldsymbol{z}}[\boldsymbol{z},\boldsymbol{z}^{\prime}]_{\xi}[\xi,t^{\prime}]_{\boldsymbol{z}^{\prime}}F^{-\beta}(t^{\prime},\boldsymbol{z}^{\prime})[t^{\prime},\infty]_{\boldsymbol{z}^{\prime}}.\nonumber 
\end{align}
Eq.~(\ref{eq:gaugeinvO}) is very close to the result of~\citep{Altinoluk:2019wyu} for the specific case of the dipole operator. With the simple
trick for any function $F$
\begin{equation}
F(\boldsymbol{z})=\int\!\frac{\rmd^{2}\boldsymbol{k}_{1}}{(2\pi)^{2}}\!\int\!\rmd^{2}\boldsymbol{b}_{1}\rme^{-i\boldsymbol{k}_{1}\cdot(\boldsymbol{b}_{1}-\boldsymbol{z})}F(\boldsymbol{b}_{1}),\label{eq:IFFtrick}
\end{equation}
introducing $\boldsymbol{r}\equiv\boldsymbol{x}_{1}-\boldsymbol{x}_{2}$
and with straightforward algebra, we can finally recover that result:
\begin{align}
\mathcal{O}_{\xi}(\boldsymbol{x}_{1},\boldsymbol{x}_{2}) & =1-ig\!\int_{\xi}^{\infty}\!\!\rmd t\int\!\!\frac{\rmd^{2}\boldsymbol{k}}{(2\pi)^{2}}\boldsymbol{r}^i\left(\frac{\rme^{i(\boldsymbol{k}\cdot\boldsymbol{x}_{1})}-\rme^{i(\boldsymbol{k}\cdot\boldsymbol{x}_{2})}}{i(\boldsymbol{k}\cdot\boldsymbol{r})}\right) \nonumber \\ & \times \!\int\!\!\rmd^{2}\boldsymbol{v}\rme^{-i(\boldsymbol{k}\cdot\boldsymbol{v})}[\infty,t]_{\boldsymbol{v}}F^{i-}(t,\boldsymbol{v})[t,\infty]_{\boldsymbol{v}}\nonumber \\
 & +(ig)^{2}\!\int_{\xi}^{\infty}\!\!\rmd t\int_{\xi}^{\infty}\!\!\rmd t^{\prime}\!\int\!\frac{\rmd^{2}\boldsymbol{k}_{1}}{(2\pi)^{2}}\!\int\!\frac{\rmd^{2}\boldsymbol{k}_{2}}{(2\pi)^{2}}\!\int\!\rmd^{2}\boldsymbol{b}_{1}\int\!\rmd^{2}\boldsymbol{b}_{2}\rme^{-i(\boldsymbol{k}_{1}\cdot\boldsymbol{b}_{1})-i(\boldsymbol{k}_{2}\cdot\boldsymbol{b}_{2})}\label{eq:DISfin}\\
 & \times\frac{\boldsymbol{r}^i \boldsymbol{r}^j}{i(\boldsymbol{k}_{2}\cdot\boldsymbol{r})}\left(\frac{\rme^{i(\boldsymbol{k}_{1}+\boldsymbol{k}_{2})\cdot\boldsymbol{x}_{1}}-\rme^{i(\boldsymbol{k}_{1}+\boldsymbol{k}_{2})\cdot\boldsymbol{x}_{2}}}{i(\boldsymbol{k}_{1}+\boldsymbol{k}_{2})\cdot\boldsymbol{r}}-\rme^{i(\boldsymbol{k}_{2}\cdot\boldsymbol{x}_{2})}\frac{\rme^{i(\boldsymbol{k}_{1}\cdot\boldsymbol{x}_{1})}-\rme^{i(\boldsymbol{k}_{1}\cdot\boldsymbol{x}_{2})}}{i(\boldsymbol{k}_{1}\cdot\boldsymbol{r})}\right)\nonumber \\
 & \times [\infty,t]_{\boldsymbol{b}_{1}}\,F^{-i}(t,\boldsymbol{b}_{1})\,[t,\xi]_{\boldsymbol{b}_{1}}\,[\boldsymbol{b}_{1},\boldsymbol{b}_{2}]_{\xi}\,[\xi,t^{\prime}]_{\boldsymbol{b}_{2}}\,F^{-j}(t^{\prime},\boldsymbol{b}_{2})\,[t^\prime,\infty]_{\boldsymbol{b}_{2}}.\nonumber 
\end{align}

\begin{figure}[H]
\begin{center}
\includegraphics[width=0.9\linewidth]{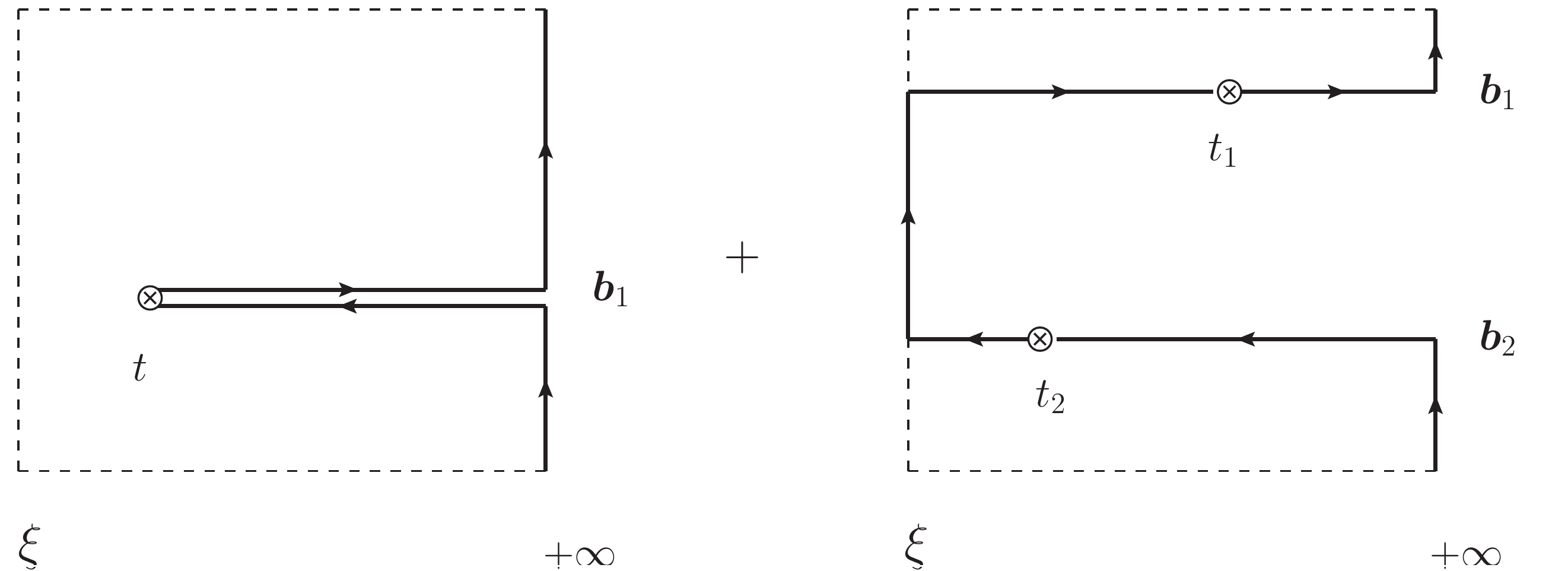}
\caption{Diagrammatic representation of the 1-body (left) and 2- body (right) amplitudes.\label{fig:one-two-amp}}
\end{center}
\end{figure}

The one-body and two-body amplitudes that appear in the above decomposition of the dipole operator are depicted in Fig.~\ref{fig:one-two-amp}.

\begin{figure}[H]
\begin{center}
\includegraphics[width=.5\linewidth]{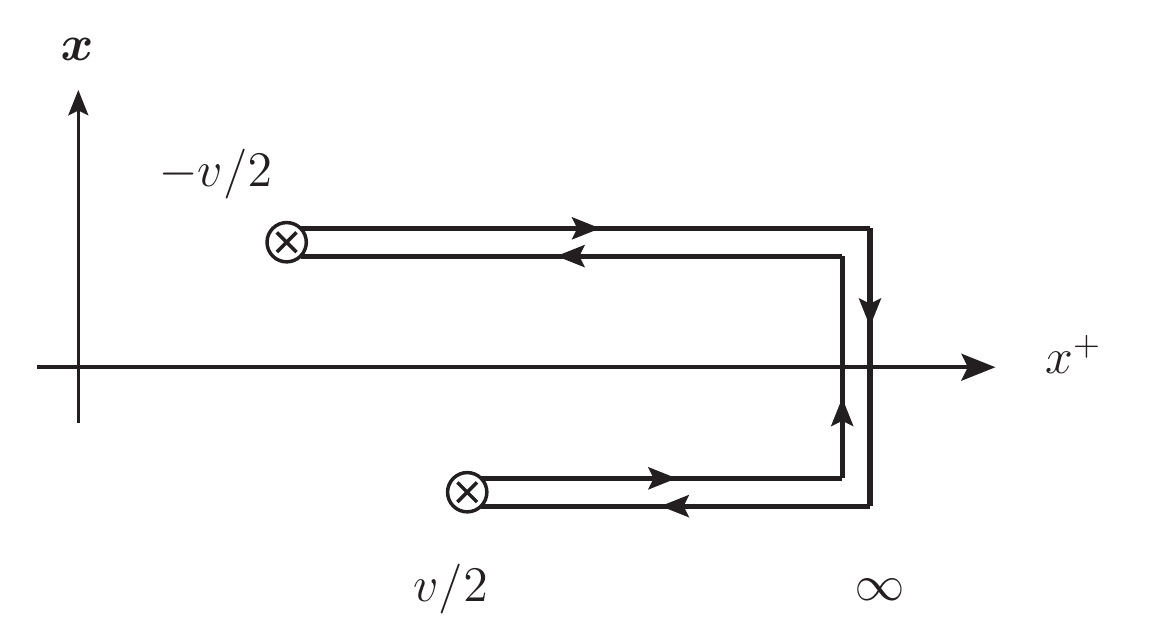}
\end{center}
\caption{Gauge link structure of the WW distribution \label{fig:WWlinks}}
\end{figure}

This result can be used in order to rewrite any small $x$ observable
involving the dipole operator in terms of TMD distributions, as in~\citep{Altinoluk:2019wyu}. Furthermore, the derivation presented above leads
to an expression for said TMD distributions which involves explicitly the previously neglected transverse
gauge links which are usually assumed to be subdominant in light cone gauge in the small
$x$ regime. Here, we found an explicitely gauge invariant expression for the operators.
\\
It is interesting to note that the derivation of Eq.~(\ref{eq:DISfin})
is straightforward once the dipole operator is fully rewritten as
the transverse string operator $\left[\hat{\boldsymbol{x}}_{1},\hat{\boldsymbol{x}}_{2}\right]_{\xi}$,
while working in the usual framework makes it quite technical. This
representation also allows for a simple understanding of the structures
encountered in~\citep{Altinoluk:2019wyu}
\begin{equation}
\frac{\rme^{i(\boldsymbol{k}\cdot\boldsymbol{r})}-1}{i(\boldsymbol{k}\cdot\boldsymbol{r})},\label{eq:struc}
\end{equation}
as the Fourier transforms of identity along the contours which support the transverse
strings. Namely,  
\beq
\int_{\x_1}^{\x_2}\rmd \z \, \,\rme^{i\z \cdot \k} = \r \,  \int_{0}^{1}\rmd s \, \,\rme^{is \r \cdot \k}\,.
\eeq
Let us consider the leading genuine twist contribution to the dipole
operator only 
\begin{align}
\mathcal{O}_{\xi}(\boldsymbol{x}_{1},\boldsymbol{x}_{2}) & =-ig\!\int_{\xi}^{\infty}\!\!\rmd t\int\!\!\frac{\rmd^{2}\boldsymbol{k}}{(2\pi)^{2}}\boldsymbol{r}^i\left(\frac{\rme^{i(\boldsymbol{k}\cdot\boldsymbol{x}_{1})}-\rme^{i(\boldsymbol{k}\cdot\boldsymbol{x}_{2})}}{i(\boldsymbol{k}\cdot\boldsymbol{r})}\right)\nonumber \\
 & \times\!\int\!\rmd^{2}\boldsymbol{v}\,\rme^{-i(\boldsymbol{k}\cdot\boldsymbol{v})}[\infty,t]_{\boldsymbol{v}}F^{i-}(t,\boldsymbol{v})[t,\infty]_{\boldsymbol{v}}\label{eq:1b}
\end{align}
and convolute it with the generic form of a $\gamma^{\left(\ast\right)}\rightarrow q\bar{q}$
hard part 
\begin{equation}
\mathcal{H}(\boldsymbol{x}_{1},\boldsymbol{x}_{2})=(2\pi)\delta(1-z-\bar{z})\rme^{-i(\boldsymbol{p}_{q}\cdot\boldsymbol{x}_{1})-i(\boldsymbol{p}_{\bar{q}}\cdot\boldsymbol{x}_{2})}\varphi(\boldsymbol{r}),\label{eq:hardpart}
\end{equation}
where we only used the constraints from longitudinal momentum conservation
through the classical field, and the Galilean boost invariance of
the photon wave function. We can easily obtain the $\mathcal{T}$
matrix for this process
\begin{align}
\mathcal{T} & =-gq^{+}\int \rmd^{2}\boldsymbol{r}\rme^{-i(\bar{z}\boldsymbol{p}_{q}-z\boldsymbol{p}_{\bar{q}})\cdot\boldsymbol{r}}\boldsymbol{r}^i \left(\frac{\rme^{i\bar{z}(\boldsymbol{p}_{q}+\boldsymbol{p}_{\bar{q}})\cdot\boldsymbol{r}}-\rme^{-iz(\boldsymbol{p}_{q}+\boldsymbol{p}_{\bar{q}})\cdot\boldsymbol{r}}}{i(\boldsymbol{p}_{q}+\boldsymbol{p}_{\bar{q}})\cdot\boldsymbol{r}}\right)\varphi(\boldsymbol{r})\nonumber \\
 & \times[\infty,0]_{\boldsymbol{0}}F^{i-}(0)[0,\infty]_{\boldsymbol{0}},\label{eq:Tm}
\end{align}
which then leads to the following cross section:
\begin{align}
\frac{d\sigma}{dzd^{2}\boldsymbol{p}_{q}d^{2}\boldsymbol{p}_{\bar{q}}} & =\int\!\!\frac{\rmd^{4}v}{(2\pi)^{3}}\delta(v^{-})\,\rme^{-i(\boldsymbol{p}_{q}+\boldsymbol{p}_{\bar{q}})\cdot\boldsymbol{v}}\left\langle P\left|F^{i-}(\frac{v}{2})\,\mathcal{U}_{\frac{v}{2},-\frac{v}{2}}^{\left[+\right]}\,F^{j-}(-\frac{v}{2})\,\mathcal{U}_{-\frac{v}{2},\frac{v}{2}}^{\left[+\right]}\,\right|P\right\rangle \nonumber \\
 & \times\frac{\alpha_{s}q^{+}}{8z\bar{z}\pi s}\int \! \rmd^{2}\boldsymbol{r}\,\rmd^{2}\boldsymbol{r}^{\prime}\rme^{-i(\bar{z}\boldsymbol{p}_{q}-z\boldsymbol{p}_{\bar{q}})\cdot\left(\boldsymbol{r}-\boldsymbol{r}^{\prime}\right)}\boldsymbol{r}^i \boldsymbol{r}^{\prime j}\,\varphi(\boldsymbol{r})\,\varphi^{\ast}(\boldsymbol{r}^{\prime})\label{eq:DISiTMD}\\
 & \times\left(\frac{\rme^{i\bar{z}(\boldsymbol{p}_{q}+\boldsymbol{p}_{\bar{q}})\cdot\boldsymbol{r}}-\rme^{-iz(\boldsymbol{p}_{q}+\boldsymbol{p}_{\bar{q}})\cdot\boldsymbol{r}}}{(\boldsymbol{p}_{q}+\boldsymbol{p}_{\bar{q}})\cdot\boldsymbol{r}}\right)\left(\frac{\rme^{-i\bar{z}(\boldsymbol{p}_{q}+\boldsymbol{p}_{\bar{q}})\cdot\left(\boldsymbol{r}-\boldsymbol{r}^{\prime}\right)}-\rme^{iz(\boldsymbol{p}_{q}+\boldsymbol{p}_{\bar{q}})\cdot\boldsymbol{r}^{\prime}}}{(\boldsymbol{p}_{q}+\boldsymbol{p}_{\bar{q}})\cdot\boldsymbol{r}^{\prime}}\right).\nonumber 
\end{align}
We clearly recognize the Weizs\"{a}cker-Williams TMD~(\ref{eq:dipdef}) from its gauge link structure, depicted in Fig.~\ref{fig:WWlinks}. 
This cross section gives the form of the cross section for observables like dijet production in DIS if it was computed with the so-called small-$x$ Improved TMD techniques~\cite{Kotko:2015ura}, thus generalizing the equivalence found in~\citep{Altinoluk:2019fui} and extended in~\citep{Altinoluk:2019wyu}. Note that the present results also contain genuine higher twist contributions, i.e. the $g^2$ terms in~(\ref{eq:DISfin}), which will not be displayed for the sake of readability. The reader is referred to~\citep{Altinoluk:2019wyu} for more explicit genuine higher twist contributions, noting that the simple and explicitely gauge invariant method we established in the present work allows for non-zero transverse gauge links in those contributions.

\section{Extension to generic color structures\label{sec:gen}}

Let us quickly extend the previous method for a more generic process:
let us consider a particle in color representation $R_{0}$ splitting
into two particles in color representations $R_{1}$ and $R_{2}$
in the external classical field. The involved Wilson line operator
is then (see e.g.~\citep{Altinoluk:2019fui}):
\begin{align}
\mathcal{O}_{\xi}^{012}(\boldsymbol{x}_{1},\boldsymbol{x}_{2}) & =\Omega_{\boldsymbol{x}_{1}}^{R_{1}}(\infty)U_{\boldsymbol{x}_{1}}^{R_{1}}(\infty,\xi)T^{R_{0}}U_{\boldsymbol{x}_{2}}^{R_{2}}(\xi,\infty)\Omega_{\boldsymbol{x}_{2}}^{R_{2}\dagger}(\infty)\label{eq:012ini}\\
 & -\Omega_{\boldsymbol{b}}^{R_{1}}(\infty)U_{\boldsymbol{b}}^{R_{1}}(\infty,\xi)T^{R_{0}}U_{\boldsymbol{b}}^{R_{2}}(\xi,\infty)\Omega_{\boldsymbol{b}}^{R_{2}\dagger}(\infty).\nonumber 
\end{align}
Here, $\boldsymbol{b}$ is the average position, weighted by longitudinal
fractions $z$ and $\bar{z}$: $\boldsymbol{b}=z\boldsymbol{x}_{1}+\bar{z}\boldsymbol{x}_{2}$. We implicitely used the following identity, for open color indices in representations $R_1$ and in $R_2$:
\begin{equation}
T^{R_0} U_{\boldsymbol{b}}^{R_0} = U_{\boldsymbol{b}}^{R_{1}}T^{R_{0}}U_{\boldsymbol{b}}^{R_{2}},\label{Fierz}
\end{equation}
\begin{figure}[H]
\begin{center}
\includegraphics[width=.45\linewidth]{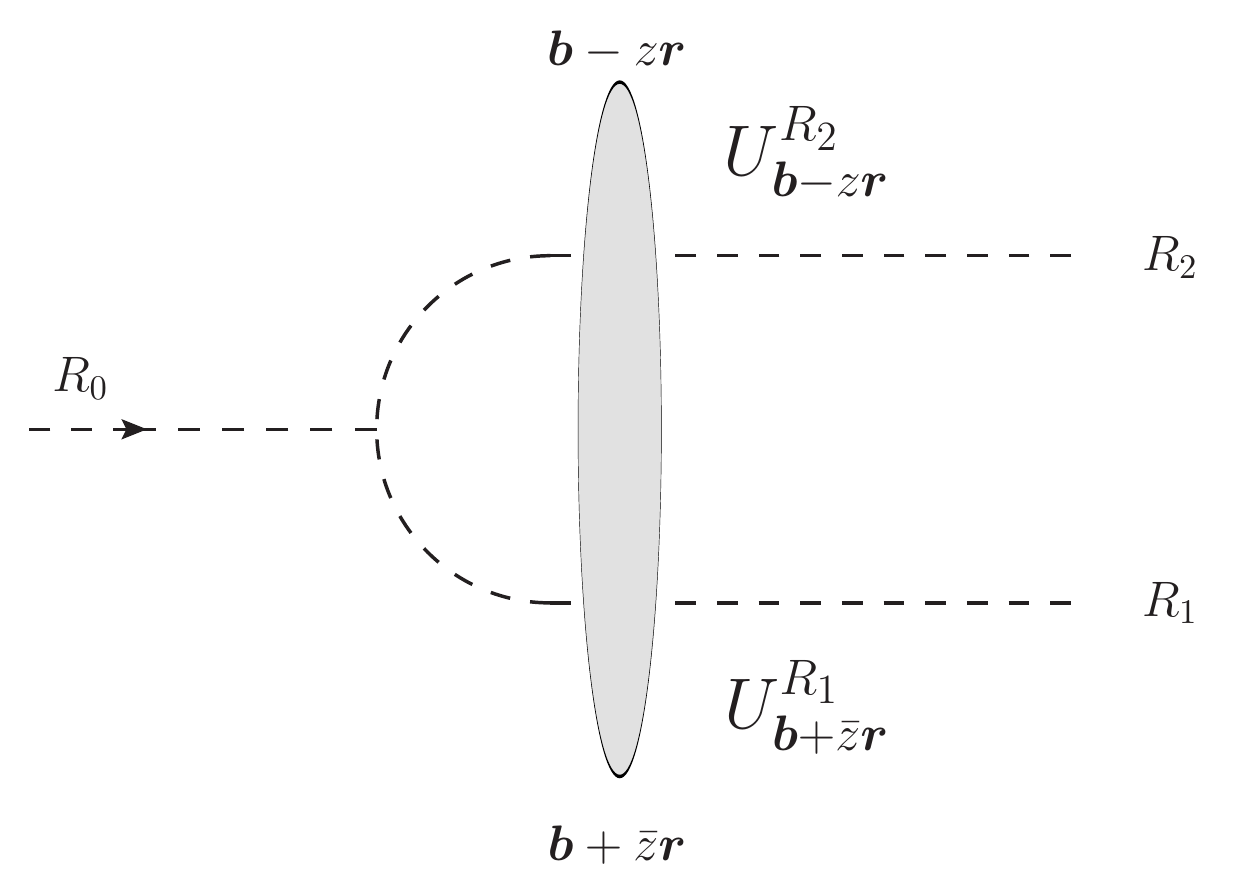} \includegraphics[width=.45\linewidth]{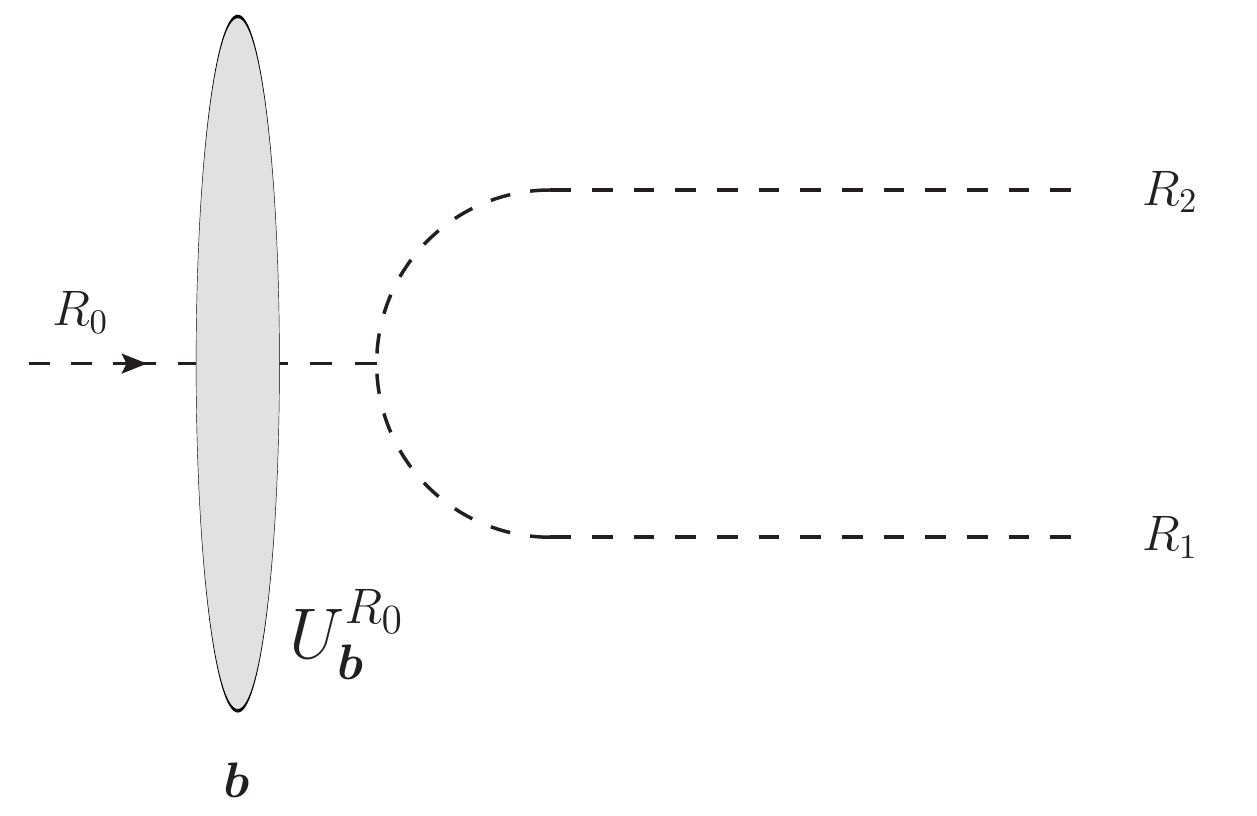}
\end{center}
\caption{Diagrams for the splitting of a particle in color representation $R_0$ into particles in respective representations $R_1$ and $R_2$, leading to the operator from Eq.~(\ref{eq:012ini})}
\end{figure}
As in the previous section, the transverse gauge links at infinity
allows to rewrite the operator with rotated Wilson lines:
\begin{align}
\mathcal{O}_{\xi}^{012}(\boldsymbol{x}_{1},\boldsymbol{x}_{2}) & \rightarrow U_{\boldsymbol{x}_{1}}^{R_{1}}(\infty,\xi)\Omega_{\boldsymbol{x}_{1}}^{R_{1}}(\xi)T^{R_{0}}\Omega_{\boldsymbol{x}_{2}}^{R_{2}\dagger}(\xi)U_{\boldsymbol{x}_{2}}^{R_{2}}(\xi,\infty)\label{eq:012-1}\\
 & -U_{\boldsymbol{b}}^{R_{1}}(\infty,\xi)\Omega_{\boldsymbol{b}}^{R_{1}}(\xi)T^{R_{0}}\Omega_{\boldsymbol{b}}^{R_{2}\dagger}(\xi)U_{\boldsymbol{b}}^{R_{2}}(\xi,\infty).\nonumber 
\end{align}
Here and from now on until the end of this chapter, we will omit the
time dependence in the intermediate equations for reader's convenience. The trick is now to write
\begin{equation}
\Omega_{\boldsymbol{x}_{1}}^{R_{1}}\,T^{R_{0}}\,\Omega_{\boldsymbol{x}_{2}}^{R_{2}\dagger}=\Omega_{\boldsymbol{x}_{1}}^{R_{1}}\,\Omega_{\boldsymbol{b}}^{R_{1}\dagger}\,\Omega_{\boldsymbol{b}}^{R_{1}}\,T^{R_{0}}\,\Omega_{\boldsymbol{b}}^{R_{2}\dagger}\,\Omega_{\boldsymbol{b}}^{R_{2}}\,\Omega_{\boldsymbol{x}_{2}}^{R_{2}\dagger},\label{eq:OmegaTrick}
\end{equation}
to interpret the $\Omega\Omega^{\dagger}$ pairs as transverse gauge
links formed from the gauge-enhanced gluon fields, see Eqs.~(\ref{eq:translink}):
\begin{equation}
\Omega_{\boldsymbol{x}_{1}}^{R_{1}}\,T^{R_{0}}\,\Omega_{\boldsymbol{x}_{2}}^{R_{2}\dagger}=[\boldsymbol{x}_{1},\boldsymbol{b}]^{R_{1}}\,\Omega_{\boldsymbol{b}}^{R_{1}}\,T^{R_{0}}\,\Omega_{\boldsymbol{b}}^{R_{2}\dagger}\,[\boldsymbol{b},\boldsymbol{x}_{2}]^{R_{2}},\label{eq:OmegaTrick2}
\end{equation}
then to absorb the Wilson lines into rotated transverse links 
\begin{equation}
U_{\boldsymbol{x}_{1}}^{R_{1}}\Omega_{\boldsymbol{x}_{1}}^{R_{1}}T^{R_{0}}\Omega_{\boldsymbol{x}_{2}}^{R_{2}\dagger}U_{\boldsymbol{x}_{2}}^{R_{2}}=[\hat{\boldsymbol{x}}_{1},\hat{\boldsymbol{b}}]^{R_{1}}\,U_{\boldsymbol{b}}^{R_{1}}\,\Omega_{\boldsymbol{b}}^{R_{1}}\,T^{R_{0}}\,\Omega_{\boldsymbol{b}}^{R_{2}\dagger}\,U_{\boldsymbol{b}}^{R_{2}}\,[\hat{\boldsymbol{b}},\hat{\boldsymbol{x}}_{2}]^{R_{2}}.\label{eq:OmegaTrick3}
\end{equation}
We can use Eqs.~(\ref{eq:linkdefL}, \ref{eq:linkdefR}) once for
each hatted link:
\begin{align}
\mathcal{O}_{\xi}^{012}(\boldsymbol{x}_{1},\boldsymbol{x}_{2}) & =-ig\!\int_{\boldsymbol{b}}^{\boldsymbol{x}_{1}}\!\!\rmd \boldsymbol{z}^i \hatA^{i \, R_{1}}(\boldsymbol{z})[\hat{\boldsymbol{z}},\hat{\boldsymbol{b}}]\,U_{\boldsymbol{b}}^{R_{1}}\,\Omega_{\boldsymbol{b}}^{R_{1}}\,T^{R_{0}}\,\Omega_{\boldsymbol{b}}^{R_{2}\dagger}\,U_{\boldsymbol{b}}^{R_{2}}\,[\hat{\boldsymbol{b}},\hat{\boldsymbol{x}}_{2}]^{R_{2}}\nonumber \\
 & -ig\!\int_{\boldsymbol{x}_{2}}^{\boldsymbol{b}}\!\!\rmd \boldsymbol{z}^iU_{\boldsymbol{b}}^{R_{1}}\,\Omega_{\boldsymbol{b}}^{R_{1}}\,T^{R_{0}}\,\Omega_{\boldsymbol{b}}^{R_{2}\dagger}\,U_{\boldsymbol{b}}^{R_{2}}\,[\hat{\boldsymbol{b}},\hat{\boldsymbol{z}}]\,\hatA^{i\,R_{2}}(\boldsymbol{z}),\label{eq:012trick}
\end{align}
or equivalently:
\begin{align}
\mathcal{O}_{\xi}^{012}(\boldsymbol{x}_{1},\boldsymbol{x}_{2}) & =-ig\!\int_{\boldsymbol{b}}^{\boldsymbol{x}_{1}}\!\!\rmd \boldsymbol{z}^i \hatA^{i\,R_{1}}(\boldsymbol{z})U_{\boldsymbol{z}}^{R_{1}}\,\Omega_{\boldsymbol{z}}^{R_{1}}\,T^{R_{0}}\,\Omega_{\boldsymbol{z}}^{R_{2}\dagger}\,U_{\boldsymbol{z}}^{R_{2}}\,[\hat{\boldsymbol{z}},\hat{\boldsymbol{x}}_{2}]^{R_{2}}\label{eq:012trick2}\\
 & -ig\!\int_{\boldsymbol{x}_{2}}^{\boldsymbol{b}}\!\!\rmd \boldsymbol{z}^{\prime j}[\hat{\boldsymbol{b}},\hat{\boldsymbol{z}}^{\prime}]^{R_{1}}\,U_{\boldsymbol{z}^{\prime}}^{R_{1}}\,\Omega_{\boldsymbol{z}^{\prime}}^{R_{1}}\,T^{R_{0}}\,\Omega_{\boldsymbol{z}^{\prime}}^{R_{2}\dagger}\,U_{\boldsymbol{z}^{\prime}}^{R_{2}}\,\hatA^{j\,R_{2}}(\boldsymbol{z}^{\prime}).\nonumber 
\end{align}
With similar tricks:
\begin{align}
\mathcal{O}_{\xi}^{012}(\boldsymbol{x}_{1},\boldsymbol{x}_{2}) & =-ig\!\int_{\boldsymbol{b}}^{\boldsymbol{x}_{1}}\!\!\rmd \boldsymbol{z}^i \boldsymbol{A}^{i\,R_{1}}(\boldsymbol{z})U_{\boldsymbol{z}}^{R_{1}}\,\Omega_{\boldsymbol{z}}^{R_{1}}\,T^{R_{0}}\,\Omega_{\boldsymbol{z}}^{R_{2}\dagger}\,U_{\boldsymbol{z}}^{R_{2}}\nonumber \\
 & -ig\!\int_{\boldsymbol{x}_{2}}^{\boldsymbol{b}}\!\!\rmd \boldsymbol{z}^iU_{\boldsymbol{z}}^{R_{1}}\,\Omega_{\boldsymbol{z}}^{R_{1}}\,T^{R_{0}}\,\Omega_{\boldsymbol{z}}^{R_{2}\dagger}\,U_{\boldsymbol{z}}^{R_{2}}\,\boldsymbol{A}^{i\,R_{2}}(\boldsymbol{z})\label{eq:012trick3}\\
 & +(ig)^{2}\!\int_{\boldsymbol{b}}^{\boldsymbol{x}_{1}}\!\!\rmd \boldsymbol{z}^i\!\int_{\boldsymbol{x}_{2}}^{\boldsymbol{z}}\!\!\rmd \boldsymbol{z}^{\prime j}\boldsymbol{A}^{i\,R_{1}}(\boldsymbol{z})U_{\boldsymbol{z}}^{R_{1}}\,\Omega_{\boldsymbol{z}}^{R_{1}}\,T^{R_{0}}\,\Omega_{\boldsymbol{z}}^{R_{2}\dagger}\,[\hat{\boldsymbol{z}},\hat{\boldsymbol{z}}^{\prime}]^{R_{2}}\,U_{\boldsymbol{z}^{\prime}}^{R_{2}}\,\boldsymbol{A}^{j\,R_{2}}(\boldsymbol{z}^{\prime})\nonumber \\
 & +(ig)^{2}\!\int_{\boldsymbol{x}_{2}}^{\boldsymbol{b}}\!\!\rmd \boldsymbol{z}^{\prime j}\!\int_{\boldsymbol{z}^{\prime}}^{\boldsymbol{b}}\!\!\rmd \boldsymbol{z}^i \boldsymbol{A}^{i\,R_{1}}(\boldsymbol{z})U_{\boldsymbol{z}}^{R_{1}}\,[\hat{\boldsymbol{z}},\hat{\boldsymbol{z}}^{\prime}]^{R_{1}}\,\Omega_{\boldsymbol{z}^{\prime}}^{R_{1}}\,T^{R_{0}}\,\Omega_{\boldsymbol{z}^{\prime}}^{R_{2}\dagger}\,U_{\boldsymbol{z}^{\prime}}^{R_{2}}\,\boldsymbol{A}^{j\,R_{2}}(\boldsymbol{z}^{\prime}).\nonumber 
\end{align}
With $\Omega_{\boldsymbol{\infty}}=1$, it is possible to replace $\Omega_{\boldsymbol{z}}^{R_{1}}(\xi)$ by $\left[\boldsymbol{z},\boldsymbol{\infty}\right]_{\xi}^{R_{1}}$
and $\Omega_{\boldsymbol{z}^{\prime}}^{R_{2}\dagger}(\xi)$ by $\left[\boldsymbol{\infty},\boldsymbol{z}^{\prime}\right]_{\xi}^{R_{2}}$.
Then using the definition of the (rotated) transverse fields, and writing the time dependence explicitely again:
\begin{align}
& \mathcal{O}_{\xi}^{012}(\boldsymbol{x}_{1},\boldsymbol{x}_{2}) \nonumber \\ & =-ig\!\int_{\xi}^{\infty}\!\!\rmd t\int_{\boldsymbol{b}}^{\boldsymbol{x}_{1}}\!\!\rmd \boldsymbol{z}^iU_{\boldsymbol{z}}^{R_{1}}(\infty,t)T_{a}^{R_{1}}F_{a}^{i-}(t,\boldsymbol{z})[t,\xi]_{\boldsymbol{z}}^{R_{1}}[\boldsymbol{z},\boldsymbol{\infty}]_{\xi}^{R_{1}}T^{R_{0}}[\boldsymbol{\infty},\boldsymbol{z}]_{\xi}^{R_{2}}U_{\boldsymbol{z}}^{R_{2}}(\xi,\infty)\nonumber \\
 & -ig\!\int_{\xi}^{\infty}\!\!\rmd t\int_{\boldsymbol{x}_{2}}^{\boldsymbol{b}}\!\!\rmd \boldsymbol{z}^iU_{\boldsymbol{z}}^{R_{1}}(\infty,\xi)[\boldsymbol{z},\boldsymbol{\infty}]_{\xi}^{R_{1}}T^{R_{0}}[\boldsymbol{\infty},\boldsymbol{z}]_{\xi}^{R_{2}}[\xi,t]_{\boldsymbol{z}}^{R_{2}}T_{a}^{R_{2}}F_{a}^{i-}(t,\boldsymbol{z})U_{\boldsymbol{z}}^{R_{2}}(t,\infty)\nonumber \\
 & +(ig)^{2}\!\int_{\xi}^{\infty}\!\!\rmd t\int_{\xi}^{\infty}\!\!\rmd t^{\prime}\!\left[\int_{\boldsymbol{b}}^{\boldsymbol{x}_{1}}\!\!\rmd \boldsymbol{z}^i \!\int_{\boldsymbol{x}_{2}}^{\boldsymbol{z}}\!\!\rmd \boldsymbol{z}^{\prime j}+\int_{\boldsymbol{x}_{2}}^{\boldsymbol{b}}\!\!\rmd \boldsymbol{z}^{\prime j}\!\int_{\boldsymbol{z}^{\prime}}^{\boldsymbol{b}}\!\!\rmd \boldsymbol{z}^i \right]\label{eq:012trick4}\\
 & \times U_{\boldsymbol{z}}^{R_{1}}(\infty,t)T_{a}^{R_{1}}F_{a}^{i-}(t,\boldsymbol{z})[t,\xi]_{\boldsymbol{z}}^{R_{1}}[\boldsymbol{z},\boldsymbol{\infty}]_{\xi}^{R_{1}}T^{R_{0}}[\boldsymbol{\infty},\boldsymbol{z}^{\prime}]_{\xi}^{R_{2}}[\xi,t^{\prime}]_{\boldsymbol{z}^{\prime}}^{R_{2}}T_{b}^{R_{2}}F_{b}^{j-}(t^{\prime},\boldsymbol{z}^{\prime})U_{\boldsymbol{z}^{\prime}}^{R_{2}}(t^{\prime},\infty).\nonumber 
\end{align}
With the use of the trick given in Eq.~(\ref{eq:IFFtrick}), one
recovers the results from~\citep{Altinoluk:2019wyu}, with explicit transverse gauge
links:
\begin{align}
& \mathcal{O}_{\xi}^{012}(\boldsymbol{x}_{1},\boldsymbol{x}_{2}) \label{eq:012fin} \\ & =-ig\bar{z}\boldsymbol{r}^i\!\int_{\xi}^{\infty}\!\!\rmd t\int\!\!\frac{\rmd^{2}\boldsymbol{k}}{(2\pi)^{2}}\!\int\!\rmd^{2}\boldsymbol{v}\rme^{-i\boldsymbol{k}\cdot(\boldsymbol{v}-\boldsymbol{x}_{1})}\frac{\rme^{i\bar{z}(\boldsymbol{k}\cdot\boldsymbol{r})}-1}{i\bar{z}(\boldsymbol{k}\cdot\boldsymbol{r})}\nonumber \\
 & \times [\infty,t]^{R_1}_{\boldsymbol{v}}T_{a}^{R_{1}}F_{a}^{i-}(t,\boldsymbol{v})[t,\xi]_{\boldsymbol{v}}^{R_{1}}[\boldsymbol{v},\boldsymbol{\infty}]_{\xi}^{R_{1}}T^{R_{0}}[\boldsymbol{\infty},\boldsymbol{v}]_{\xi}^{R_{2}}[\xi,\infty]_{\boldsymbol{v}}^{R_{2}}\nonumber \\
 & -igz\boldsymbol{r}^i\!\int_{\xi}^{\infty}\!\!\rmd t\int\!\frac{\rmd^{2}\boldsymbol{k}}{(2\pi)^{2}}\!\int\!\rmd^{2}\boldsymbol{v}\,\rme^{-i\boldsymbol{k}\cdot(\boldsymbol{v}-\boldsymbol{x}_{2})}\frac{\rme^{iz\left(\boldsymbol{k}\cdot\boldsymbol{r}\right)}-1}{iz(\boldsymbol{k}\cdot\boldsymbol{r})}\nonumber \\
 & \times [\infty,\xi]^{R_1}_{\boldsymbol{v}}[\boldsymbol{v},\boldsymbol{\infty}]_{\xi}^{R_{1}}T^{R_{0}}[\boldsymbol{\infty},\boldsymbol{v}]_{\xi}^{R_{2}}[\xi,t]_{\boldsymbol{v}}^{R_{2}}T_{a}^{R_{2}}F_{a}^{i-}(t,\boldsymbol{v})[t,\infty]_{\boldsymbol{v}}^{R_{2}}\nonumber \\
 & +(ig)^{2}\!\int_{\xi}^{\infty}\!\!\rmd t\int_{\xi}^{\infty}\!\!\rmd t^{\prime}\!\int\!\!\frac{\rmd^{2}\boldsymbol{k}_{1}}{(2\pi)^{2}}\!\int\!\!\frac{\rmd^{2}\boldsymbol{k}_{2}}{(2\pi)^{2}}\!\int\!\rmd^{2}\boldsymbol{b}_{1}\int\!\rmd^{2}\boldsymbol{b}_{2}\,\rme^{-i\boldsymbol{k}_{1}\cdot(\boldsymbol{b}_{1}-\boldsymbol{b})}\rme^{-i\boldsymbol{k}_{2}\cdot(\boldsymbol{b}_{2}-\boldsymbol{b})}\nonumber \\
 & \times\frac{\boldsymbol{r}^i \boldsymbol{r}^j}{(\boldsymbol{k}_{1}+\boldsymbol{k}_{2})\cdot\boldsymbol{r}}\left(\frac{\rme^{-i(\boldsymbol{k}_{2}\cdot\boldsymbol{r})}-1}{(\boldsymbol{k}_{2}\cdot\boldsymbol{r})}\rme^{i\bar{z}(\boldsymbol{k}_{1}+\boldsymbol{k}_{2})\cdot\boldsymbol{r}}+\frac{\rme^{i(\boldsymbol{k}_{1}\cdot\boldsymbol{r})}-1}{(\boldsymbol{k}_{1}\cdot\boldsymbol{r})}\rme^{-iz(\boldsymbol{k}_{1}+\boldsymbol{k}_{2})\cdot\boldsymbol{r}}\right)\nonumber \\
 & \times [\infty,t]_{\boldsymbol{b}_{1}}^{R_{1}}T_{a}^{R_{1}}F_{a}^{i-}(t,\boldsymbol{b}_{1})[t,\xi]_{\boldsymbol{b}_{1}}^{R_{1}}[\boldsymbol{b}_{1},\boldsymbol{\infty}]_{\xi}^{R_{1}}T^{R_{0}}[\boldsymbol{\infty},\boldsymbol{b}_{2}]_{\xi}^{R_{2}}[\xi,t^{\prime}]_{\boldsymbol{b}_{2}}^{R_{2}}T_{b}^{R_{2}}F_{b}^{j-}(t^{\prime},\boldsymbol{b}_{2})[t^{\prime},\infty]_{\boldsymbol{b}_{2}}^{R_{2}}.\nonumber 
\end{align}

\section{Extension to 3-line operators\label{sec:3L}}

The formulation of Wilson line operators in terms of transverse gauge
links makes it very easy to extend the proof of equivalence to TMD
distributions for observables with more than 2 Wilson lines involved.
For example, let us consider the $\gamma^{\left(\ast\right)}\rightarrow q\bar{q}g$
amplitude. It involves operators with 2 Wilson lines, which can be
treated as in previous sections, but it also contains a 3-line operator\footnote{For simplicity, we use the standard small $x$ formulation where the
Wilson lines are extended to $\xi=-\infty$: $U_{\boldsymbol{x}}^{R}=\left[\infty,-\infty\right]_{\boldsymbol{x}}^{R}.$}
\begin{equation}
\mathcal{M}_{3}=U_{\boldsymbol{x}_{1}}\bt^{b}W_{\boldsymbol{x}_{3}}^{ba}U_{\boldsymbol{x}_{2}}^{\dagger},\label{eq:o3}
\end{equation}
which in the fundamental representation reads as a 4-line one:
\begin{equation}
\mathcal{M}_{3}=U_{\boldsymbol{x}_{1}}U_{\boldsymbol{x}_{3}}^{\dagger}\bt^{a}U_{\boldsymbol{x}_{3}}U_{\boldsymbol{x}_{2}}^{\dagger}.\label{eq:o3F}
\end{equation}
Taking into account all 3 factors $\Omega(\infty)$ and using the tricks
from Section~\ref{sec:par-transporter}, this operator becomes
\begin{equation}
\mathcal{M}_{3}=\left[\hat{\boldsymbol{x}}_{1},\hat{\boldsymbol{x}}_{3}\right]_{\xi}\bt^{a}\left[\hat{\boldsymbol{x}}_{3},\hat{\boldsymbol{x}}_{2}\right]_{\xi}.\label{eq:o3trans}
\end{equation}
Applying Eq.~(\ref{eq:trick}) for $\xi=-\infty$ simultaneously
on the left and on the right of the color matrix and using the explicit
expressions for the rotated fields~(\ref{eq:AiF}) allows to rewrite
directly $\mathcal{U}_{3}$ with 1-, 2-, 3- and 4-body contributions,
reading respectively\footnote{Note that there is a non-scattering contribution as well in the definition
of the operator in Eq.~(\ref{eq:o3}), which we subtracted.}:
\begin{align}
\mathcal{M}_{3}^{\left(1\right)} & =-ig\int\!\rmd t\int_{\boldsymbol{x}_{3}}^{\boldsymbol{x}_{1}}\rmd \boldsymbol{z}^i[\infty,t]_{\boldsymbol{z}}F^{i-}(t,\boldsymbol{z})[t,\infty]_{\boldsymbol{z}}\bt^{b}\nonumber \\
 & -ig\int\!\rmd t\int_{\boldsymbol{x}_{2}}^{\boldsymbol{x}_{3}}\rmd \boldsymbol{z}^i \bt^{b}[\infty,t]_{\boldsymbol{z}}F^{i-}(t,\boldsymbol{z})[t,\infty]_{\boldsymbol{z}},\label{eq:O31}
\end{align}
\begin{align}
\mathcal{U}_{3}^{\left(2\right)} & =\left(ig\right)^{2}\int\!\rmd t\!\int\!\rmd t^{\prime}\!\int_{\boldsymbol{x}_{3}}^{\boldsymbol{x}_{1}}\!\rmd \boldsymbol{z}^i\!\int_{\boldsymbol{x}_{3}}^{\boldsymbol{z}}\!\rmd \boldsymbol{z}^{\prime j} \nonumber \\
 & \times [\infty,t]_{\boldsymbol{z}}F^{i-}(t,\boldsymbol{z})\,[t,\xi]_{\boldsymbol{z}}[\boldsymbol{z},\boldsymbol{z}^{\prime}]_{\xi}[\xi,t^{\prime}]_{\boldsymbol{z}^{\prime}}F^{j-}(t^{\prime},\boldsymbol{z}^{\prime})\,[t^\prime,\infty]_{\boldsymbol{z}^\prime}\bt^{b}\nonumber \\
 & +\left(ig\right)^{2}\int\!\rmd t\!\int\!\rmd t^{\prime}\!\int_{\boldsymbol{x}_{3}}^{\boldsymbol{x}_{1}}\!\rmd \boldsymbol{z}^i \!\int_{\boldsymbol{x}_{2}}^{\boldsymbol{x}_{3}}\!\rmd \boldsymbol{z}^{\prime j} \nonumber \\ 
 & \times [\infty,t]_{\boldsymbol{z}}F^{i-}(t,\boldsymbol{z})\,[t,\infty]_{\boldsymbol{z}}\,\bt^{b}\,[\infty,t^\prime]_{\boldsymbol{z}^{\prime}}F^{j-}(t^{\prime},\boldsymbol{z}^{\prime})\,[t^\prime,\infty]_{\boldsymbol{z}^\prime}\nonumber \\
 & +\left(ig\right)^{2}\int\!\rmd t\!\int\!\rmd t^{\prime}\!\int_{\boldsymbol{x}_{2}}^{\boldsymbol{x}_{3}}\!\rmd \boldsymbol{z}^i \!\int_{\boldsymbol{x}_{2}}^{\boldsymbol{z}}\!\rmd \boldsymbol{z}^{\prime j} \nonumber \\ 
 & \times \bt^{b}\,[\infty,t]_{\boldsymbol{z}}F^{i-}(t,\boldsymbol{z})\,[t,\xi]_{\boldsymbol{z}}[\boldsymbol{z},\boldsymbol{z}^{\prime}]_{\xi}[\xi,t^{\prime}]_{\boldsymbol{z}^{\prime}}F^{j-}(t^{\prime},\boldsymbol{z}^{\prime})\,[t^\prime,\infty]_{\boldsymbol{z}^\prime},\label{eq:O32}
\end{align}
\begin{align}
\mathcal{U}_{3}^{\left(3\right)} & =-\left(ig\right)^{3}\int\!\rmd t_{1}\int\!\rmd t_{2}\int\!\rmd t_{3}\int_{\boldsymbol{x}_{3}}^{\boldsymbol{x}_{1}}\!\rmd \boldsymbol{z}^i \int_{\boldsymbol{x}_{3}}^{\boldsymbol{z}}\rmd \boldsymbol{z}^{\prime j}\int_{\boldsymbol{x}_{2}}^{\boldsymbol{x}_{3}}\rmd \boldsymbol{u}^k \nonumber \\
 & \times [\infty,t_1]_{\boldsymbol{z}}F^{i-}(t_{1},\boldsymbol{z})[t_{1},\xi]_{\boldsymbol{z}}[\boldsymbol{z},\boldsymbol{z}^{\prime}]_{\xi}[\xi,t_{2}]_{\boldsymbol{z}^{\prime}}F^{j-}(t_{2},\boldsymbol{z}^{\prime}) \nonumber \\ 
 & \times [t_2,\infty]_{\boldsymbol{z}^{\prime}}\bt^{b}[\infty,t_3]_{\boldsymbol{u}}F^{k-}(t_{3},\boldsymbol{u})[t_3,\infty]_{\boldsymbol{u}}\nonumber \\
 & -\left(ig\right)^{3}\int\!\rmd t_{1}\int\!\rmd t_{2}\int\!\rmd t_{3}\int_{\boldsymbol{x}_{3}}^{\boldsymbol{x}_{1}}\!\rmd \boldsymbol{z}^i \int_{\boldsymbol{x}_{2}}^{\boldsymbol{x}_{3}}\!\rmd \boldsymbol{u}^j \int_{\boldsymbol{x}_{2}}^{\boldsymbol{u}}\!\rmd \boldsymbol{u}^{\prime k}\label{eq:O33}\\
 & \times [\infty,t_1]_{\boldsymbol{z}}F^{i-}(t_{1},\boldsymbol{z})[t_1,\infty]_{\boldsymbol{z}}\bt^{b}[\infty,t_2]_{\boldsymbol{u}}F^{j-}(t_{2},\boldsymbol{u}) \nonumber \\ 
 & \times [t_{2},\xi]_{\boldsymbol{u}}[\boldsymbol{u},\boldsymbol{u}^\prime]_{\xi}[\xi,t_{3}]_{\boldsymbol{u}^\prime}F^{k-}(t_{3},\boldsymbol{u}^\prime)[t_3,\infty]_{\boldsymbol{u}^\prime},\nonumber 
\end{align}
and
\begin{align}
\mathcal{U}_{3}^{\left(4\right)} & =\left(ig\right)^{4}\int\!\!d t_{1}\!\int\!\rmd t_{2}\!\int\!\rmd t_{3}\!\int\!\rmd t_{4}\!\int_{\boldsymbol{x}_{3}}^{\boldsymbol{x}_{1}}\!\rmd \boldsymbol{z}^i\int_{\boldsymbol{x}_{3}}^{\boldsymbol{z}}\!\rmd \boldsymbol{z}^{\prime j} \int_{\boldsymbol{x}_{2}}^{\boldsymbol{x}_{3}}\! \rmd \boldsymbol{u}^k \int_{\boldsymbol{x}_{2}}^{\boldsymbol{u}}\!\rmd \boldsymbol{u}^{\prime l}\nonumber \\
 & \times [\infty,t_1]_{\boldsymbol{z}}F^{i-}(t_{1},\boldsymbol{z})\,[t_{1},\xi]_{\boldsymbol{z}}[\boldsymbol{z},\boldsymbol{z}^{\prime}]_{\xi}[\xi,t_{2}]_{\boldsymbol{z}^{\prime}}F^{j-}(t_{2},\boldsymbol{z}^{\prime})\,[t_2,\infty]_{\boldsymbol{z}^{\prime}}\,\bt^{b}\nonumber \\
 & \times [\infty,t_3]_{\boldsymbol{u}}F^{k-}(t_{3},\boldsymbol{u})\,[t_{3},\xi]_{\boldsymbol{u}}[\boldsymbol{u},\boldsymbol{u}^\prime]_{\xi}[\xi,t_{4}]_{\boldsymbol{u}^\prime}F^{l-}(t_{4},\boldsymbol{u}^\prime)\,[t_4,\infty]_{\boldsymbol{u}^\prime}\label{eq:O34}
\end{align}
Although the final expressions are quite cumbersome, the method
to derive them is straightforward once the dipoles have been replaced
by transverse strings. All it took was Eq.~(\ref{eq:trick}). 

\section{Discussions}
We have provided a reinterpretation of operators built from Wilson lines as operators built from transverse strings. This observation allows for compact and systematic extensions of the exact small-$x$/TMD equivalence shown in~\citep{Altinoluk:2019wyu}. It also allows for a systematic power expansion which preserves QCD gauge invariance at each step. We have given an example of extension beyond previously established results. \\
Besides the fact that small $x$ models can lead to interesting insight on TMD distributions at asymptotic energies, the infinite-twist TMD form of small $x$ amplitudes also changes our understanding of saturation effects in the dilute-dense regime~\citep{Altinoluk:2019wyu} by distinguishing between kinematic and genuine saturation effects. In light of this new development, it would be interesting to revisit the non-linear terms in small $x$ evolution in terms of TMD distributions. The first step towards the understanding of these non-linearities is precisely what we accomplished in our example from Section~\ref{sec:3L}, by extracting the TMD operators from the 3-line operator which appears when one evolves a dipole. \\
Another advantage of the TMD form of small $x$ amplitudes is the opportunity it opens for the use of TMD evolution equations to resum logarithms of the hard scale as well as Sudakov logarithms. Thanks to the extension described in Section~\ref{sec:3L}, it is now possible to adapt the full Next-to-Leading Logarithmic cross section for inclusive DIS~\citep{Beuf:2017bpd,Hanninen:2017ddy} in order to include these resummations.

\section*{Acknowledgements} 
This work is supported by the U.S. Department of Energy, Office of Science, Office of Nuclear Physics, under contract No. DE- SC0012704, and in part by Laboratory Directed Research and Development (LDRD) funds from Brookhaven Science Associates.


\appendix

\bibliographystyle{elsarticle-num}

\bibliography{MasterBibtex}

\end{document}